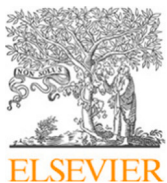

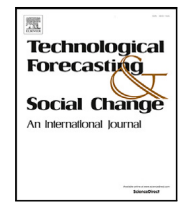

# Projections of Earth's technosphere: Scenario modeling, worldbuilding, and overview of remotely detectable technosignatures

Jacob Haqq-Misra [a][,][*], George Profitiliotis [a], Ravi Kopparapu [b]

[a] *Blue Marble Space Institute of Science, 600 1st Avenue, 1st Floor, Seattle, WA, 98104, USA*
[b] *NASA Goddard Space Flight Center, 8800 Greenbelt Road, Greenbelt, MD, 20771, USA*

## ARTICLE INFO

## ABSTRACT

This study uses methods from futures studies to develop a set of ten self-consistent scenarios for Earth's 1000-year future, which can serve as examples for defining technosignature search strategies. We apply a novel worldbuilding pipeline that evaluates the dimensions of human needs in each scenario as a basis for defining the observable properties of the technosphere. Our scenarios include three with zero-growth stability, two that have collapsed into a stable state, one that oscillates between growth and collapse, and four that continue to grow. Only one scenario includes rapid growth that could lead to interstellar expansion. We examine absorption spectral features for a few scenarios to illustrate that nitrogen dioxide can serve as a technosignature to distinguish between present-day Earth, pre-agricultural Earth, and an industrial 1000-year future Earth. Three of our scenarios are spectrally indistinguishable from pre-agricultural Earth, even though these scenarios include expansive technospheres. Up to nine of these scenarios could represent steady-state examples that could persist for much longer timescales, and it remains possible that short-duration technospheres could be the most abundant. Our scenario set provides the basis for further systematic thinking about technosignature detection as well as for imagining a broad range of possibilities for Earth's future.

## 1. Introduction

Astrobiology seeks to understand the "origin, distribution, and future of life in the universe" (Space Studies Board, 2019), which includes the use of ground- and space-based astronomical observatories to search for possible biosignatures or technosignatures that would be indicative of extraterrestrial life (e.g., Fujii et al., 2018; Meadows et al., 2023). The search for spectroscopic biosignatures in exoplanet atmospheres has been motivated by theoretical exploration of changes in Earth's spectral signature through time (e.g., Arney et al., 2016, 2018; Schwieterman et al., 2018) as well as observational studies of Earth's spectral signature today (e.g., Robinson et al., 2011; Sterzik et al., 2012). Understanding the historical evolution of Earth's biosphere and associated spectral signature has also inspired theoretical exploration of alternative exoplanetary biosignatures (e.g., Krissansen-Totten et al., 2016, 2018) and false positives for biosignature detection (e.g., Catling et al., 2018; Meadows et al., 2018; Harman and Domagal-Goldman, 2018; Foote et al., 2023). The actual discovery of an exoplanetary biosignature may be unlike any of the historical or theoretical possibilities that have so far been explored, but the only way to develop the requisite technology and search strategy for something as unknown as extraterrestrial life is to begin with the known example of Earth's present and past biosphere.

Technology is a relatively recent development on Earth, and the search for technosignatures in exoplanetary systems can only draw upon the recent past and present-day conditions as a known basis for motivating actual searches. Any theoretical understanding of future changes in Earth's biosphere and technosphere must therefore draw upon present-day conditions to make projections of plausible future scenarios. This approach has routinely been invoked by the technosignature (or SETI, search for extraterrestrial intelligence) research community, which engages in speculation about the detectability of technology that does not yet exist but plausibly could exist (e.g., Haqq-Misra et al., 2022c; Socas-Navarro et al., 2021). For example, the Ĝ infrared search for extraterrestrial civilizations (Wright et al., 2014a,b; Griffith et al., 2015; Wright et al., 2015) was an extensive analysis of data from mid-infrared surveys to look for possible infrared excesses that could be evidence of technological megastructures (i.e., Dyson spheres or swarms; human civilization has not yet built any Dyson sphere elements, but the theoretical possibility of such megastructures (e.g., Dyson, 1960; Wright, 2023) makes them viable as plausible future technology on Earth—or existing technology elsewhere. Other studies (e.g., Kopparapu et al., 2021; Haqq-Misra et al., 2022a,b) have considered the detectability of pollutants in exoplanet atmospheres at abundances many times greater than on Earth today; such futures may be less optimistic, but the theoretical possibility of elevated pollution on future Earth suggests the plausibility of such technosignatures in exoplanet atmospheres. Radio SETI, optical SETI, and numerous other examples in technosignature research all follow this approach of informally making projections (often linear or exponential) about the future, which serves as a basis for performing an assessment of detectability or designing the specifications of an instrument.

Perhaps the most classic example of this approach is the scale that was developed by Kardashev (1964) for describing "technologically developed civilizations" as they expand through space. We focus specifically on the Kardashev scale because this idea has become foundational in the technosignature research community (see, e.g., Sagan, 1973; Carrigan, 2012; Cirkovic, 2015; Besteiro, 2019; Gray, 2020; Wright et al., 2022) and is also frequently invoked in other research that seeks to project possible trajectories of human civilization (see, e.g., Baker, 2020; Namboodiripad and Nimal, 2021; Jiang et al., 2022; Zhang et al., 2023). The underlying assumption of Kardashev (1964) is that the growth in power consumed by human civilization — and perhaps a technological civilization in general — will continue to the point where expansion into space becomes necessary:

Assuming [a] 1% [growth rate], we find that the energy consumption per second will be equal to the output of the sun per second, 3200 years from now, i.e., $4 \times 10^{22}$ erg/s, and that in 5800 years the energy consumed will equal the output of $10^{11}$ stars like the sun. The figures arrived at seem to be inordinately high when compared to the present level of development, but we see no reasons why the tempo of increase in energy consumption should fall substantially than predicted (Kardashev, 1964, p.218).

Following this logic, Kardashev (1964) defined a three-tier scale for describing civilizational development:

I — technological level close to the level presently attained on the earth, with energy consumption at $\approx 4 \times 10^{19}$ erg/s.

II — a civilization capable of harnessing the energy radiated by its own star (for example, the stage of successful construction of a "Dyson sphere"); energy consumption at $\approx 4 \times 10^{33}$ erg/s.

III — a civilization in possession of energy on the scale of its own galaxy, with energy consumption at $\approx 4 \times 10^{44}$ erg/s (Kardashev, 1964, p.219).

This model of continuous growth for thousands of years into the future has been used to motivate observational and theoretical approaches to technosignature science by searching for civilizations that might follow such trajectories. For example, Sagan (1973) suggested that "the best policy might therefore be to search with existing technology for Type II or Type III civilizations among the nearer galaxies, rather than Type I or younger civilizations among the nearer stars". The analysis by Wright et al. (2022) similarly asserted that "technosignatures, by contrast [to other biosignatures], have essentially no upper limit to their detectability", with a reference to the Kardashev (1964) scale. Others have critiqued the idea of continuous growth and suggested that any long-lived civilizations might instead expand at a slower rate, or not expand into space at all (e.g., Von Hoerner, 1975; Newman and Sagan, 1981; Haqq-Misra and Baum, 2009; Mullan and Haqq-Misra, 2019; Likavčan, 2024). Such arguments, and their rebuttals, have all drawn upon informal methods that contain implicit assumptions, which inevitably biases any results toward what is familiar and known.

The interdisciplinary field of futures studies has developed a variety of systematic methodological approaches toward developing self-consistent future trajectories. The field uses the plural "futures" to indicate that it does not attempt to *predict* a singular future but instead is an approach to *project* multiple different futures that illustrate a range of possible scenarios. Such methods attempt to avoid both technological determinism and social determinism when discussing future technological developments; instead, human technology and human society are seen as intertwined, interacting, co-evolving, and capable of mutually shaping each other. Rather than making reductionist assumptions about continual rates of growth, futures studies methods draw upon insights from social sciences and other disciplines to understand the intertwined influences of human cultures with their political, economic, and social systems.

Futures studies methods are widely used in a diverse range of applications, which includes strategic planning by government, military, medical, and commercial organizations. For example, the Royal Dutch/Shell Group has gained a reputation as a pioneer in the use of scenario development as a tool for corporate long-range planning (e.g., Cornelius et al., 2005; Jefferson, 2012), which has been mimicked and extended by other organizations that need to entertain multiple possible future trajectories to account for deep uncertainty. Other applications include the development of emissions scenarios for use in the Intergovernmental Panel on Climate Change (IPCC) reports, which enable climate modeling groups to explore the response of the Earth system to various scenarios of future greenhouse gas forcing up to the year 2100. The use of scenario development does not assert that any one of the scenarios explored will necessarily be correct, but instead scenario planning is intended to provide a set of self-consistent plausible cases to illuminate the range of possibilities and develop an appropriate set of responses that could perform robustly.

Technosignature science can benefit from a more systematized approach toward envisioning possible futures, with trajectories of Earth's futures serving as self-consistent scenarios for the assessment of technosignature detectability and search strategy. The use of futures studies methods has been suggested by Voros (2018) as a relevant tool for mapping the morphology of extraterrestrial contact scenarios, while Profitiliotis and Theologou (2023) have suggested that training in futures studies methods and ways of thinking (known as "futures literacy") could help to prepare for the unknown dimensions of discovering extraterrestrial life, including intelligent life. The use of scenarios has also been suggested by Ramirez et al. (2015) as a rigorous methodology for generating "interesting research" that can enable scholars to meaningfully engage with complex and uncertain problems. In this study, we use scenarios of Earth's future as templates for possible extraterrestrial technology. The greatest difficulty in applying futures studies methods to the projection of Earth's future technosignatures is the timescale: existing methods in futures studies make projections on short timescales of years to decades, but no generalized methods exist for envisioning futures on much longer century to millennium timescales. The development of detailed projections across long time scales is difficult, but the problem becomes more tractable when the goal is to predict changes in Earth's technosphere that could be detectable through astronomical observation or exploration.

This paper represents the first attempt to develop a systematic approach toward the construction of long-term future scenarios of Earth's technosphere. We focus on a 1000-year future trajectory for the scenarios in this study, which is longer than the timescale considered by existing IPCC emissions scenarios and long enough that planetary-scale transformations such as terraforming could conceivably occur. The methods used in this study may be unfamiliar to astrobiologists, astronomers, and other readers, and our application of existing methods may also be novel to practitioners in futures studies. We introduce and describe each method in context as it is used in order to allow the widest range of readers to follow our approach. The methodological approach of the paper is divided into three parts. In Section 2 we develop a large set of scenarios, analyze them for consistency, and combine those that are similar to obtain a final set of ten long-term scenarios for Earth's future. In Section 3 we construct a pipeline for developing the details for each scenario that are most salient for defining the observable properties of the technosphere. In Section 4 we provide an overview of the potentially detectable technosignatures across the range of scenarios. The paper also includes a discussion (Section 5) of the most immediate implications for the search for technosignatures. However, the information generated by this scenario development study is rich, and complete exploration of the scenarios and their implications for astrobiology will follow in subsequent papers.

It is again worth emphasizing at the outset that this approach is intended to explore the plausibility space for Earth's future technosphere. The ten scenarios that are generated and analyzed in this paper do not necessarily represent assertions about what the actual future will be like, nor do any of these ten scenarios represent the assessment of the authors or the method itself as to the likelihood or probability of any of these futures. Instead, the purpose of this study is to develop trajectories for Earth's future that can enable the technosignature community to expand its thinking beyond traditional assumptions of linear growth to imagine a much broader set of possibilities.

## 2. Scenario modeling

Part one of our method is a process for defining a finite set of scenarios from the total set of all possible futures. This process begins by defining the "scenario space" (i.e., the full set of possible future scenarios) in terms of several key parameters of interest, which can help to systematically explore as many possibilities as we can. This approach follows a method known as "general morphological analysis", which is a general-purpose approach for the stepwise development of scenarios (Johansen, 2018). The set of scenarios generated by this method is intended to minimize the bias in underlying assumptions and span a wide range of possibilities.

The general morphological analysis has three steps, which we briefly describe before demonstrating the process. In step one, we first state the problem in as exact a formulation as possible, and then break the problem down into a number of parameters that are "meaningful, equally important, abstract, straightforward, independent of each other, and have many internal connections" (Johansen, 2018). In step two, we then use these parameters to construct a multidimensional matrix that describes the full scenario space, where each unique combination of parameter values defines a scenario. In step three, we perform a cross-consistency assessment to eliminate any inconsistent scenarios (i.e., those that are logically impossible or paradoxical), which will result in a final set of plausible and self-consistent scenarios for further analysis. We note that our use of the term "self-consistent" refers to our use of a cross-consistency assessment (described further below) to eliminate scenarios that would contain self-contradictory, impossible, or unlikely qualities, so that our final set of scenarios are only those that represent plausible and realizable futures.

Here we state the problem as: "What are the technological phenomena of the future anthrosphere and how can they be described?" The phrasing of this question draws upon Earth system science (Steffen et al., 2020), with the anthrosphere as part of the larger Earth system. The technosphere is included as a subset of the anthrosphere, along with other subsystems such as economy, institutions, and cultures. We can consider the technosphere as the material manifestation of infrastructure that is shaped by the immaterial subsystems of the anthrosphere. This statement of the problem enables our analysis to focus on understanding the technological phenomena that emerge from the plausible set of future anthrospheres. It is worth noting that our analysis focuses only on anthrospheres that are viable and functional as continuities when "observed", which makes them relevant and interesting under a principle of mediocrity. In other words, we do not investigate anthrospheres that have completely collapsed to the point of human extinction or are on the verge of particular turning points that are effectively special occasions of discontinuity in the duration of 1000-year civilizational timescales. We acknowledge the possibility that "graveyard technosignatures" (see, e.g., Haqq-Misra and Baum, 2009; Villarroel et al., 2022) that serve as lingering evidence of a now-extinct technological civilization are a possible subject of interest and even could be the most prominent technosignature in the galaxy or universe. However, including graveyard systems is beyond the scope of our analysis, which is based on the coupling between human and technological systems. We acknowledge that further extensions of our methods could be applied to graveyard scenarios, but we leave such considerations for a subsequent study.

We break down the problem using the widely applied PEST (political, economic, social, and technological) framework (Aguilar, 1967), which considers four factors that can provide meaningful and important descriptions of future scenarios. We note that many variations of the PEST framework exist, which can include explicit consideration of environmental factors as well as legal factors (such as the PESTEL framework). Our approach does not neglect environmental (or legal) implications, but we do not prescribe these factors during our scenario modeling. Instead, environmental (and relevant legal) factors are all emergent properties of our scenarios, which result from the political, economic, social, and technological factors that define each scenario. Our interest in this study is specifically in understanding the possibility space for Earth's future technosphere, so we further decompose our parameter space into two sets: one focusing on political, economic, and social factors (Section 2.1) and another focusing specifically on technology factors (Section 2.2). For each of these two sets, we develop a multidimensional matrix based on a fixed number of parameters, each of which can be assigned to one of several possible values in an exhaustive way. We then assess each matrix for cross-consistency and develop a final set of scenarios based on the remaining sets of factors from the two matrices. We note at the outset that our approach is intended to capture the interdependencies that exist between political, economic, social, and technological factors, rather than assuming that any of these is necessarily determined exclusively by others. Our use of a second matrix for technology factors alone does not at all imply that technology is determined by political, economic, or social factors (or vice versa) and instead should be interpreted as a bidirectional relationship of co-shaping. As will be seen in the approach that follows, our separate selection of global factors and technological factors is to enable an assessment of bidirectional consistency between these two sets of factors, rather than implying a deterministic dependency of one on the other.

### 2.1. Global factors

We start the general morphological analysis by focusing on the first three factors of the PEST framework: the political, economic, and social factors. We use these three factors to define our first morphological matrix, which describes future human society in terms of its global institutions. The purpose of this study is to characterize detectable properties of planetary technospheres, so our assignment of political, economic, and social factors will be representative of global trends taken in aggregate. This approach does not account for any local or regional variations at this stage, but such heterogeneities will be included later during the worldbuilding phase (Section 3). We construct the global factors morphological matrix in Table 1 with these three parameters to describe the global economic, political, and social systems for our scenario space. We selected the particular values for each of these factors deductively based on our need to be both generalizable and comprehensive in the construction of our matrix of the immaterial part of the anthrosphere. A *generalized* set of values for these parameters is necessary for making projections on a 1000-year time horizon because the set of present-day political, economic, and social ideologies and philosophies will not necessarily reflect the set of such ideologies that are present in any given future; rather than employing horizon scanning to identify trends, uncertainties, and weak signals that may have inductively helped map out multiple alternatives for the evolution of current ideologies over time horizons of two to three decades, we have adopted a more generalized set of values that can capture the essence of both existing ideologies and any new possibilities that might emerge later in the course of our 1000-year time horizon. A *comprehensive* set of values refers to our choice of the most fundamental units of variability that can logically occur in the possibility space for each parameter; such choices are intended to construct our matrix in a way that includes all permutations of how political, social, and economic factors could change in the most fundamental ways. These factors are further described in this subsection.

**Table 1**
Global factors multidimensional matrix. This matrix specifies the possible values for the economic, political, and social system factors across our scenario space.

| X | Economy | Y | Politics | C | Society |
|---|---------|---|----------|---|---------|
| X1 | Scarcity | Y1 | Rule by one | Z1 | Hierarchical |
| X2 | Non-scarcity | Y2 | Rule by few | Z2 | Distributed |
| | | Y3 | Rule by all | | |
| | | Y4 | Rule by none | | |

The first factor is the global economic system (X). This factor addresses the issue of producing and distributing scarce resources among groups/communities of humans. Specific solutions for economic systems are numerous, but we attempt to avoid restricting our long-term future projections to any specific known solutions. We instead consider the possibilities that the future economy will operate either under scarcity (X1, most resources are limited) or achieve non-scarcity (X2, most resources are unlimited). A scarcity economy is consistent with present-day Earth, in which nearly all commodities are limited and prices can vary based on availability. A non-scarcity economy, by contrast, would have most or all resources available as unlimited, especially for basic human survival needs, in the way that breathable air is generally not considered as a commodity on Earth today. These values for the economy factor are admittedly a simplification of the range of existing economic ideologies, but the use of a binary selection between scarcity and non-scarcity provides a generalizable approach to grossly categorize current and future ideologies based on their most fundamental essence; likewise, these choices allow for a comprehensive description of future economic systems, as the choice between scarcity and non-scarcity possibilities covers the full range of possibilities.

The second factor is the global political system (Y). This factor addresses the issue of rule or sovereignty over the collective lives of groups, states, or other associations of humans. The choice of values for this factor should also avoid the tendency to rely on any existing known political theories and instead should take general form that can be extended to long-term futures. We consider four options for the global political system to capture a complete set of possibilities. Rule by one (Y1) involves a single actor with a complete hold of global political power, rule by few (Y2) involves a small number of actors that hold the majority of global political power, rule by all (Y3) involves global political power being widely disseminated, and rule by none (Y4) involves the absence of or a breakdown in the organization of global politics. The Y2 factor is the most consistent with present-day Earth. As with the economy factor, the values chosen for this political factor are also intended to be generalizable beyond present-day political ideologies; likewise, the selection of four values provides a comprehensive description of the possible configurations of political systems at the most fundamental unit of variability.

The third factor is the global social system (Z), which addresses the issue of how groups, states, or other associations of humans are organized. As with the other factors, the values for this factor should take general form to be applicable to long-term scenarios. We select two possible options for the global social system, which can be either organization in hierarchical structures (Z1) in which top-down social dynamics dominate or distributed structures (Z2) in which horizontal social interactions are primary. The Z1 factor is the most consistent with present-day Earth. The values chosen for the social factor are abstracted in the most generalizable way, to reflect either a vertical or horizontal, i.e., egalitarian, distribution of power as the most prominent feature; this binary selection between hierarchical and distributed systems likewise is comprehensive, with the assumption that any present-day or future social system will be fundamentally characterized by one of these two choices.

The completed global factors multidimensional matrix (Table 1) consists of three factors that give $2 \times 4 \times 2 = 16$ different solutions. These solutions all describe unique possibilities for future scenarios; however, not all of these solutions may be viable or meaningful. For example, the idea of political rule by one (Y1) is inconsistent with the idea of distributed social structures (Z2), so any solutions with the Y1-Z2 pair can be eliminated. We therefore proceed to the next step in the general morphological analysis and conduct a cross-consistency assessment, where we systematically eliminate any inconsistent value pairs to arrive at a final set of global factors scenarios.

The general approach to a cross-consistency assessment is to inspect each pair of values in the multidimensional matrix to determine whether a contradiction or inconsistency exists. There are twenty unique possible combinations of value pairs in Table 1. For this study, we use Anthropic's "Claude,"[1] a large language model (LLM) that has been trained using the "Constitutional AI" process (Bai et al., 2022), which sets it apart from other LLMs. This process trained Claude to be helpful, honest, and harmless, not via human feedback labels for harmful outputs but via self-improvement steered by a small set of principles that comprise a "constitution" that governs its behavior. Furthermore, Artificial Intelligence-assisted scenario development based on LLMs is already being explored within the futures studies community and has been described as "promising" for the generation of base material to be considered further by human experts in a hybrid strategic foresight process (Spaniol and Rowland, 2023). Combining the systematized set of modeled scenarios constructed via general morphological analysis with the generative strengths of Claude, which was found by the authors to be competent in scenario development methods, presents an excellent opportunity for producing an initial substrate for evaluation and modification by human experts. It should be noted that we only utilized the LLM as a descriptive tool, prompting it to express the multiple parameter value combinations in natural language and to describe in natural language the ramifications of particular value pairs being present together. We harness this opportunity by invoking the LLM to conduct a first cross-consistency assessment of the global factors value pairs with explicit concise arguments for and against cross-consistency, which we then review manually and correct as needed. (We note that the use of the LLM saved significant time but still resulted in some erroneous assessments. The assessment was facilitated by the LLM's arguments, but was eventually conducted by humans, which is the standard practice in general morphological analysis.) We use a 4-point Likert scale to evaluate each value pair's consistency: 0 for no consistency, 1 for low consistency, 2 for moderate consistency, and 3 for high consistency. For our analysis of global factors, we retain value pairs of moderate and high consistency for the next steps of the analysis, while the rest are rejected. The results of this cross-consistency assessment are shown in Fig. 1, where the cells with a score of 2 or 3 indicate consistent value pairs and those with a score less than 2 have been deemed inconsistent. This gives a cross-consistency matrix in which eight of the twenty possible value pairs have been eliminated.

With the cross-consistency assessment complete, the remaining value pairs in the global factors cross-consistency matrix (Fig. 1) result in three viable scenarios, which are summarized in Table 2. To better understand the features and differences provided by each of these combinations, we prompt the Claude LLM to express in natural language the value combinations of these three global factors scenarios for a thousand years into the future, generating illustrative descriptions.

---

[1] https://www.anthropic.com/claude.

**Table 2**

Summary of global factors scenarios. The selection of economy, politics, and society values is from the cross-consistency analysis (Fig. 1). The archetype and myth/metaphor are descriptions for understanding the qualities and trajectory of each scenario.

| Scenario | Economy | Politics | Society | Archetype | Myth/Metaphor |
|---|---|---|---|---|---|
| GF1 | X1 (Scarcity) | Y1 (Rule by one) | Z1 (Hierarchical) | Discipline/Collapse | Philosopher-king |
| GF2 | X1 (Scarcity) | Y2 (Rule by few ) | Z1 (Hierarchical) | Growth/Collapse | Survival of the Fittest |
| GF3 | X2 (Non-scarcity) | Y3 (Rule by all ) | Z2 (Distributed) | Transformation | Pure Democracy |

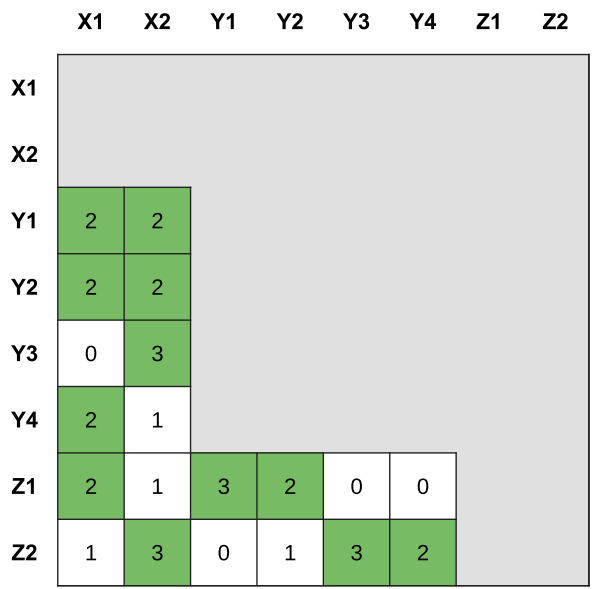

**Fig. 1.** Global factors cross-consistency assessment. The 12 green cells with scores of 2 or 3 indicate consistent value pairs. The remaining 8 cells with scores less than 2 indicate inconsistent value pairs. (For interpretation of the references to color in this figure legend, the reader is referred to the web version of this article.)

The generated text for each scenario provides an initial basis as an entry point for our deeper (human) analysis of the themes underlying each scenario. Without assistance from the LLM, we deductively assign an "archetype" to each scenario based on the four generic images of the future system created by Dator (see, e.g., Dator, 2009; Bezold, 2009): in Dator's general model, any given scenario can be characterized as undergoing a state of growth (upward trajectory), a state of collapse (downward trajectory), a state of discipline (steady or horizontal trajectory), or a state of transformation (non-linear or disjointed trajectory). These archetypes help us frame subsequent worldbuilding options after the combination of these intermediate scenarios with the technology factors intermediate scenarios into the complete, final, nuanced anthroposphere scenarios. Next, again without assistance from the LLM, we inductively assign a "myth/metaphor" for each scenario to reflect its guiding civilizational undercurrent, which draws upon a method known as "causal layered analysis" (Inayatullah, 1998) applied to understand the deepest underlying motivations of the actors and systems in a given scenario. This inductive human analysis facilitates better making sense of the illustrative descriptions generated by the LLM, as well as their critical evaluation and the identification of the most salient way of describing the themes that emerge from each scenario.

Scenario GF1 involves the combination X1-Y1-Z1, which is a scarcity economy in which one actor rules a hierarchical social structure. We characterize this as a "discipline/collapse" archetype to indicate the increasing concentration of power and control in the hands of a single ruler and the decreasing quality of life for most people. We describe GF1 as the myth of the "philosopher-king", which indicates the prominence of a single autocrat's decisions over political and social spheres in this scenario.

Scenario GF2 involves the combination X1-Y2-Z1, which is a scarcity economy in which a few actors rule over hierarchical social structures. This scenario is most similar to present-day Earth. We characterize this as a "growth/collapse" archetype to indicate that this is a scenario in which both growth and collapse trajectories have occurred at some point from the present day to the projected thousand-year future. We describe GF2 as the myth of "survival of the fittest", which indicates the prominence of unbridled competitive forces over the political and social systems in this scenario.

Scenario GF3 involves the combination X2-Y3-Z2, which is a non-scarcity economy in which political power and social structures are distributed. We characterize this as a "transformation" archetype to indicate that this is a scenario in which breakthrough developments have occurred to enable new modes of non-scarcity economics and distributed organization that would not otherwise have been possible. We describe GF3 as the myth of "pure democracy", which indicates the prominence of distributed decision-making and resource-sharing in this scenario.

The construction of the global factors scenarios is now complete. It is worth noting that none of the scenarios include value Y4 (rule by none), as all possible combinations with this value were eliminated during the cross-consistency assessment. The result of this effort is the reduction of our set of global factors scenarios from a total of sixteen possibilities to three that have been identified as internally consistent for a thousand year future projection. These three global factors scenarios will be combined later with our set of scenarios for future technology, which we turn to next.

### 2.2. Technology factors

We continue the general morphological analysis by constructing a second matrix that focuses on the technology factor remaining from the PEST framework. The purpose of constructing a separate morphological matrix for technology factors alone is to gain finer detail for describing the possible developments of a future human technosphere, which will be useful in thinking about the range of possible technosignatures in future Earth scenarios. This approach assesses the state of technology as a whole across the domain of the anthroposphere and neglects any local variations, although regional heterogeneities will be considered later during the worldbuilding phase (Section 3). We construct the technology morphological matrix with five factors (Table 3), which are intended to capture a broad set of possibilities for the trajectory of human technology. These factors are described further in this subsection.

The first factor is the technosphere's relationship to the biosphere (A). This factor addresses the extent to which the technosphere and biosphere overlap with one another. Today the biosphere and technosphere have some elements in common: Earth has a biosphere and a technosphere that interact in some ways, while the technosphere has expanded across other parts of the solar system from space activities. We can describe the possibilities for the relationship between these Earth system spheres by drawing upon set theory to give five unique values: the biosphere and technosphere are equivalent sets (A1), the biosphere is a proper subset of the technosphere (A2), the technosphere is a proper subset of the biosphere (A3), the biosphere and the technosphere have some common elements while not being equal (A4), and the biosphere and the technosphere have no common elements (A5). The A4 value is consistent with present-day Earth.

The second factor is the spatial distribution of the majority of the technosphere's technomass (B). The vast majority of the technosphere's technomass in our solar system today is spatially concentrated around one central location or "pole", planet Earth, and this factor considers the future distribution of this technomass in space. This factor assumes that Earth is a point in space and considers the whole of space as open for possible distributions of technology, which gives four unique values: unipolar (B1), bipolar (B2), multipolar (B3), and non-polar (B4). The B1 value is consistent with present-day Earth.

**Table 3**
Technology factors multidimensional matrix. This matrix specifies the possible values for five factors describing the technosphere across our scenario space.

| A | Relationship to biosphere | B | Spatial distribution | C | Development | D | Highest order | E | Smallest scale |
|---|---|---|---|---|---|---|---|---|---|
| A1 | The Biosphere and Technosphere are equivalent sets | B1 | Unipolar | C1 | Evolved and emergent | D1 | First-Order (technologies that relate humans to the natural world) | E1 | Less than 1 nm |
| A2 | The Biosphere is a proper subset of the Technosphere | B2 | Bipolar | C2 | Designed and directed | D2 | Second-Order (technologies that relate humans to other technologies) | E2 | From 1 nm to 1 μm |
| A3 | The Technosphere is a proper subset of the Biosphere | B3 | Multipolar (more than 2) | | | D3 | Third-Order (technologies that relate technologies to other technologies) | E3 | More than 1 μm |
| A4 | The Biosphere and Technosphere have some common elements while not being equal | B4 | Non-polar | | | | | | |
| A5 | The Biosphere and Technosphere have no common elements | | | | | | | | |

The third factor is the nature or intent of the technosphere's development (C). The development of the technosphere on Earth today is the result of emerging technologies that lead to competition and improvements, but we do not centrally plan the development of the technosphere on a planetary level. This factor considers the nature of possible future developments of the technosphere, which could include the contrasting possibility of central planning and construction of a technosphere. The two values for this factor describe the development of the technosphere as either evolved and emergent (C1), or designed and directed (C2). The C1 factor is consistent with present-day Earth.

The fourth factor is the highest order of technology reached in the technosphere (D). This factor addresses the extent to which technological elements mediate interactions between agents and realms, physical and symbolic, from a teleological perspective. This factor considers three values for describing the highest extent of technology-mediated relationships achieved in a future technosphere: first-order relationships (D1) are technologies with the purpose of helping humans to interact with the physical world; second-order relationships (D2) are technologies whose purpose is to help humans interact with an otherwise inaccessible symbolic realm; and third-order relationships (D3) are technologies whose purpose is to help completely autonomous technological agents interact with the physical and symbolic realms without a human in the loop. For present-day Earth, the highest order of technologies reached in the technosphere is that of second-order technologies (the D2 value), but the other values for this parameter describe possible highest orders in a future technosphere.

The final factor is the smallest scale of interconnected and interdependent systems in the technosphere (E). The smallest interconnected and interdependent systems that are part of the technosphere today are marginally at the nanometer scale, whereas early civilizations on Earth had a much larger and macroscopic scale at which technological systems connected. This parameter considers the smallest scale of such technology networks in a future technosphere, which takes three possible values: less than 1 nm (E1), from 1 nm to 1 μm (E2), and more than 1 μm (E3). The E2 parameter is consistent with present-day Earth.

The completed technology factors multidimensional matrix (Table 3) consists of five parameters that give $5 \times 4 \times 2 \times 3 \times 3 = 360$ different solutions. We again note that not all of these possible scenarios may be viable, as some combinations of value pairs may be inconsistent or illogical. The solution space for our technology factors is also fairly large, so any reduction in the set of total technology scenarios will make subsequent analysis less burdensome.

As we did in Section 2.1, we conduct a cross-consistency assessment for the value pairs in the technology factors matrix to eliminate any pairs that would be inconsistent. The use of the LLM is even more helpful in expressing value pairs in natural language and providing explicit concise arguments for and against cross-consistency for this second matrix, as there are 113 unique possible combinations of value pairs in Table 3; however, the reasoning and assessment are tasks for the human researchers, despite the LLM's natural language assistance, so we still need to manually inspect and evaluate all the initial results and rationales and make corrections. We again use a 4-point Likert scale to evaluate each value pair's consistency; for our analysis of technology factors, we apply a more strict criterion and only maintain pairs of high consistency (with a score of 3) for propagation to the next steps of the analysis, while the remaining value pairs are rejected. The resulting technology cross-consistency matrix eliminates 50 of the 113 possible value pairs as inconsistent or unlikely, as shown in Fig. 2.

With the cross-consistency assessment complete, the remaining value pairs in the technology factors cross-consistency matrix (Fig. 2) result in eleven viable scenarios, which are summarized in Table 4. To better imagine and comprehend these value combinations, we again prompt the Claude LLM to express in natural language the value combinations of these eleven technology factors scenarios for a thousand years into the future, generating illustrative descriptions. The LLM scenario descriptions suggested that some scenarios show significant similarities to the extent that they could be grouped together. We choose to manually cluster the technology factors scenarios in an effort to further reduce the total number required for analysis. Afterwards, leveraging causal layered analysis as before, we inductively assign a myth/metaphor for each Cluster, again without assistance from the LLM.

The first cluster of technology factors (Cluster 1) includes TF6, TF7, TF9, and TF10. These combinations differ only on the spatial distribution of the technosphere being bipolar (B2) versus multipolar (B3) as well as whether the highest possible technology interactions are second-order (D2) or third-order (D3). We describe Cluster 1 using the myth/metaphor of "living parallel to machines", which indicates a designed technosphere managed by automation of partial or complete autonomy. The second cluster (Cluster 2) includes TF8 and TF11. These

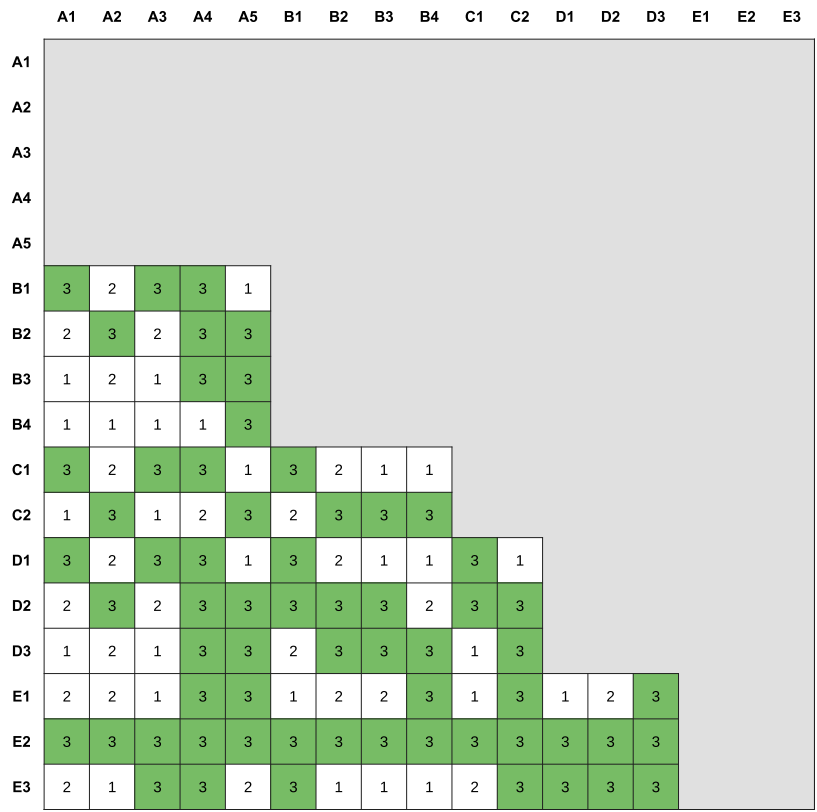

**Fig. 2.** Technology factors cross-consistency assessment. The 63 green cells with scores of 3 indicate consistent value pairs. The remaining 50 cells with scores less than 3 indicate inconsistent value pairs.

**Table 4**

Summary of technology factors scenarios. The selection of values is from the cross-consistency analysis (Fig. 2). The myth/metaphor is a description for understanding the qualities and trajectory of each scenario. Scenarios with identical myths will be clustered.

| Scenario | Relationship | Distribution | Development | Order | Scale | Myth/Metaphor |
|---|---|---|---|---|---|---|
| TF1 | A1 (Equivalent sets) | B1 (Unipolar) | C1 (Evolved) | D1 (1st-order) | E2 (nm to μm) | Environmental Sustainability |
| TF2 | A2 (Bio subset of tech) | B2 (Bipolar) | C2 (Designed) | D2 (2nd-order) | E2 (nm to μm) | Planetary Engineering |
| TF3 | A3 (Tech subset of bio) | B1 (Unipolar) | C1 (Evolved) | D1 (1st-order) | E2 (nm to μm) | Environmental Sustainability |
| TF4 | A4 (Some common elements) | B1 (Unipolar) | C1 (Evolved) | D1 (1st-order) | E2 (nm to μm) | Simplicity |
| TF5 | A4 (Some common elements) | B1 (Unipolar) | C1 (Evolved) | D2 (2nd-order) | E2 (nm to μm) | Earth 2024 |
| TF6 | A5 (No common elements) | B2 (Bipolar) | C2 (Designed) | D2 (2nd-order) | E2 (nm to μm) | Living Parallel to Machines |
| TF7 | A5 (No common elements) | B2 (Bipolar) | C2 (Designed) | D3 (3rd-order) | E2 (nm to μm) | Living Parallel to Machines |
| TF8 | A5 (No common elements) | B4 (Non-polar) | C2 (Designed) | D3 (3rd-order) | E1 (< nm) | Technology Flees Earth |
| TF9 | A5 (No common elements) | B3 (Multipolar) | C2 (Designed) | D2 (2nd-order) | E2 (nm to μm) | Living Parallel to Machines |
| TF10 | A5 (No common elements) | B3 (Multipolar) | C2 (Designed) | D3 (3rd-order) | E2 (nm to μm) | Living Parallel to Machines |
| TF11 | A5 (No common elements) | B4 (Non-polar) | C2 (Designed) | D3 (3rd-order) | E2 (nm to μm) | Technology Flees Earth |

combinations differ only on whether the smallest level of technological network achieved is at the micro (E2) versus nano (E1) scale. We describe the myth of Cluster 2 as "technology flees Earth", which indicates a technosphere that has emigrated to other parts of the solar system and beyond. Both Cluster 1 and Cluster 2 have biology and technology as separately directed entities, as indicated by the A5-C2 pair. Cluster 1 involves technosphere and biosphere co-existing on Earth, while Cluster 2 involves cases of technosphere panspermia that leave Earth's biosphere at an almost pre-technological state.

The third cluster (Cluster 3) includes TF1 and TF3, which differ only on whether the technosphere and biosphere are equivalent sets (A1) or if all technology is part of the biosphere (A3). We describe the myth of Cluster 3 as "environmental sustainability", which indicates a technosphere that is significantly engulfed by the biosphere. The fourth cluster (Cluster 4) contains TF2. We describe Cluster 4 as "planetary engineering" to indicate a technosphere that has completely transformed planetary systems. The fifth cluster (Cluster 5) contains TF4. We describe Cluster 5 as "simplicity" to indicate a technosphere in which much of present-day technology has been lost. The final sixth cluster (Cluster 6) contains TF5, which we describe as "Earth 2024" because it corresponds to a present-day technosphere.

The construction of the technology factors scenarios is now complete. It is worth noting that none of the scenarios include the E3 factor, in which the smallest scale of interacting systems in the technosphere are larger than a micrometer. In other words, the E3 factor represents a technosphere with only macroscopic interacting elements, which corresponds approximately to a "steampunk" scenario that leverages simple technologies to build a complex society. However, such imaginative possibilities did not remain after the cross-consistency assessment. We next complete our scenario modeling by combining the six technology factors scenario clusters with the global factors scenario clusters.

### 2.3. Final scenarios

We began this scenario modeling by considering a scenario space that included 16 global factors scenarios × 360 technology factors scenarios = 5760 possible scenarios. This total scenarios space is intractable for analysis and also contains many cases that are inconsistent, and our cross-consistency assessments reduced this total scenario space to a much smaller number of 3 global factors scenarios × 11 technology factors scenarios = 33 unique scenarios. We also identified six clusters in the technology factors scenarios, which further reduces the total number of unique scenarios to 18. This is a much more manageable number of scenarios for analysis, but we continue to further assess the compatibility between the combinations of global and technology factors in an effort to further reduce the final number of scenarios.

| | TF1 | TF2 | TF3 | TF4 | TF5 | TF6 | TF7 | TF8 | TF9 | TF10 | TF11 |
|-----|-----|-----|-----|-----|-----|-----|-----|-----|-----|------|------|
| **GF1** | N | N | N | N | Y | N | N | N | N | N | N |
| **GF2** | Y | Y | Y | Y | Y | N | N | N | N | N | N |
| **GF3** | Y | Y | Y | N | Y | Y | Y | Y | Y | Y | Y |

**Fig. 3.** Compatibility assessment between global factors scenarios and technology factors scenarios. The 16 green cells marked 'Y' indicate consistent parings. The remaining 17 cells marked 'N' indicate inconsistent pairings. (For interpretation of the references to color in this figure legend, the reader is referred to the web version of this article.)

We perform a compatibility assessment of the 33 viable global and technology scenarios to identify any combinations of global factors and technology factors parts that are inconsistent on the basis of their corresponding myths/metaphors that reflect their deeper civilizational foundations. We take an interactional theory stance (Friedman and Hendry, 2019) on the immaterial and material manifestations of the future anthroposphere, captured by the global and technology factors: we assume that they must be mutually shaped and compatible from a civilizational perspective at their deepest cultural level, that is, at the level of collective metaphors and myths. As before, we use the Claude LLM only to generate illustrative descriptions of the potential combinations of global factors and technology factors parts to help us in our imagining of the complexity of those anthroposphere combinations, but the assessment itself is conducted manually by the human researchers explicitly adopting an interactional theory stance. As discussed previously, the criterion used for this assessment is the compatibility of the deeper civilizational undercurrents of the global factors and technology factors parts of each candidate scenario as co-evolving and co-shaping aspects of the anthroposphere, especially as reflected in the level of their underlying myths/metaphors. This part of the assessment assigns only binary options for compatibility (yes/no) and does not include any additional scoring. The results of the compatibility assessment are shown in Fig. 3, which leaves 16 remaining combinations between the global and technology factors that are compatible at the myth/metaphor level of their underpinning civilizations.

When we then consider the clusters of technology factors, this 16 combinations further reduces to a total of just 10 scenarios for Earth's 1000 year technosphere. These final scenarios are summarized in Table 5, with each scenario now assigned an identifier ranging from S1–S10. The final nuanced underlying myth/metaphor for each scenario will be discussed further in Section 3.2. A visualization of the final scenario space is also shown in Fig. 4, which shows all the remaining combinations of global and technology factors. We have now reduced the total scenario space of 5760 possibilities by 0.17% to obtain these ten final scenarios. This completes part one of our method, and we will proceed to further examine these ten scenarios in the next section.

## 3. Worldbuilding

Part two of our method is a process for systematically constructing details for each scenario in order to arrive at self-consistent descriptions of the technosphere. This process is known as "worldbuilding", which in general has a wide range of applications from military or corporate strategic planning to the development of fictional worlds for film or video games. We first provide a brief overview of the role of worldbuilding in futures studies, especially in relation to scenarios, which, however, tend to focus on shorter timescales of decades in the future. We then describe our novel pipeline for constructing long-term projections of Earth's technosphere and then apply this pipeline to our set of ten scenarios.

**Table 5**
Summary of final scenarios. Global factors are from Table 2 and technology factors are from Table 4. The final myth/metaphor is a nuanced description for understanding the qualities and trajectory of each complete scenario.

| Scenario | Global factors | Technology factors | Myth/Metaphor |
|----------|----------------|--------------------|--------------------|
| S1 | GF1 | Cluster 6 (TF5) | Big Brother is Watching |
| S2 | GF2 | Cluster 6 (TF5) | Wild West |
| S3 | GF3 | Cluster 6 (TF5) | Golden Age |
| S4 | GF2 | Cluster 5 (TF4) | Living with the Land |
| S5 | GF3 | Cluster 4 (TF2) | Transhumanism |
| S6 | GF2 | Cluster 4 (TF2) | Sword of Damocles |
| S7 | GF3 | Cluster 3 (TF1, TF3) | Restoration |
| S8 | GF2 | Cluster 3 (TF1, TF3) | Ouroboros |
| S9 | GF3 | Cluster 2 (TF8, TF11) | Deus Ex Machina |
| S10 | GF3 | Cluster 1 (TF6, TF7, TF9, TF10) | Out of Eden |

Worldbuilding is the process of creating "imaginary worlds with coherent geographic, social, cultural, and other features" (Von Stackelberg and McDowell, 2015). A key added benefit brought by worldbuilding to futures studies is its ability to produce persistent coherent worlds situated in future times and places to immerse audiences in future scenarios. In turn, this deeper engagement with the future can be used both to better understand the underlying layers of scenarios and to reassess the present from a new lens, leading to novel insights. Worldbuilding has been proposed as a tool for creating "thick descriptions" of possible future worlds, that is, dense and detailed images of the future that allow for further exploration of the future and reflection, both on the desirability of a future and on the present worldviews and assumptions that are left unquestioned (Mehnert, 2021). These logically consistent descriptions are particularly valuable in capturing the entanglement and co-evolution of humans and technology as socio-technical systems embedded in their cultural contexts.

### 3.1. Pipeline

Each of our ten scenarios has a unique set of global factors and technology factors, as summarized in Table 5. These unique combinations all correspond to different possible trajectories of the future; however, it is difficult based on Table 5 to imagine the specific detectable technosignatures that could arise in each of these scenarios. For this reason we turn to the process of worldbuilding, which enables us to begin with each set of global and technology factors in order to develop the details of each world so that we can describe the properties of the technosphere. Existing worldbuilding methods are insufficient for describing 1000-year futures, so we develop our own worldbuilding approach that combines elements from existing methodologies into a novel "pipeline" for describing each future technosphere.

A diagram of our worldbuilding "pipeline" is shown in Fig. 5. This pipeline begins with the unique inputs for each scenario and ends with recommendations for technosignature detection, with the intermediate steps used to self-consistently develop the relevant details in each scenario. Each step of the worldbuilding pipeline is described further in the text that follows. We note that completing this pipeline for a given scenario involves some subjectivity and creativity on our part, which is a feature of any worldbuilding process. Attempts by other investigators to follow this pipeline for the same set of scenario inputs may lead to different descriptions of each future world than ours, which may even lead to different manifestations of the technosphere. This should not be considered as a methodological weakness, but instead this illustrates the uncertainties in making long-term projections and serves to motivate further use of this pipeline to transparently map rationales on the basis of prior premises. Our use of this pipeline in this study results in ten self-consistent technospheres that can be used for further analysis, and any use of this pipeline by others will also result in a self-consistent and diverse set of future technospheres that span the range of global and technological factors. This pipeline may also be relevant to other problems in futures studies, but such applications are beyond the scope of this paper.

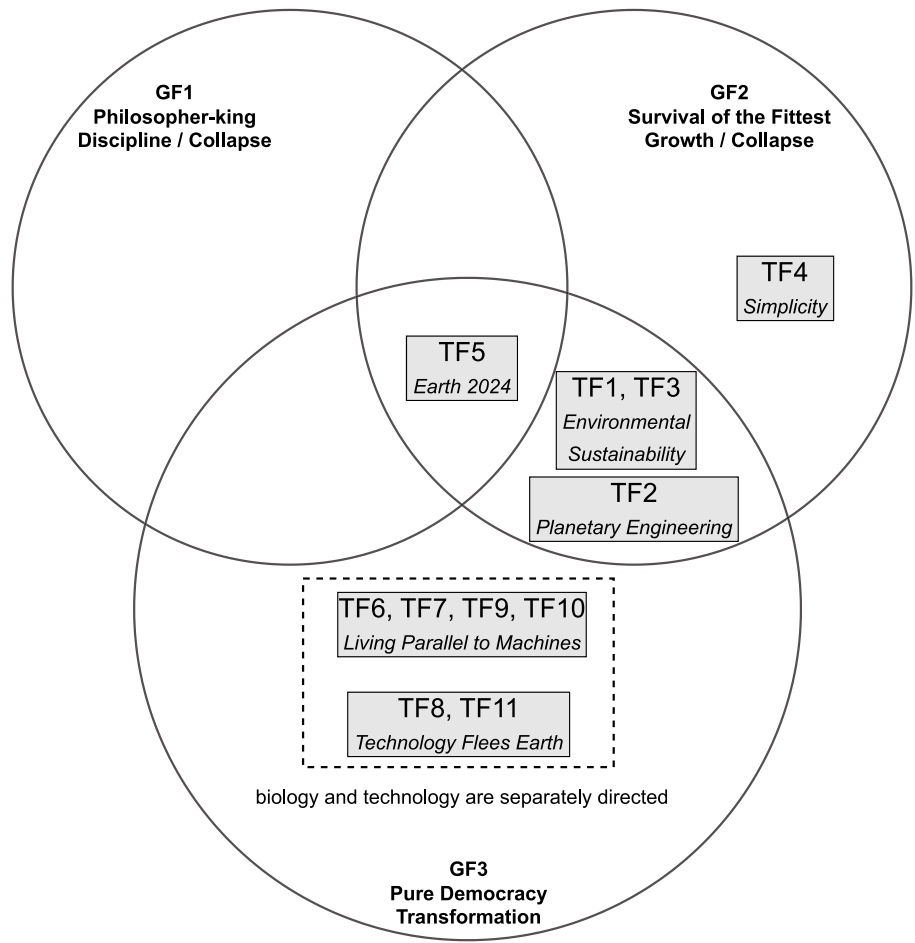

**Fig. 4.** Venn diagram showing the association of six technology factor scenario clusters with the three global factor scenarios.

The worldbuilding pipeline begins by selecting a particular scenario and stating the basic assumptions (Fig. 5, gray boxes). Each scenario has a single global factors scenario and a technology factors scenario cluster (Table 5). The pipeline also includes a set of basic assumptions that are common to all scenarios: (1) humans have not gone extinct; (2) humans have not speciated; (3) humans are the only terrestrial animal capable of producing technology; (4) no extraterrestrial technology has interfered with human technological development; and (5) the scenario takes place 1000 years in the future. All content in the pipeline must remain consistent with these global and technology factors as well as the basic assumptions.

In the next step of the pipeline, we generate a scenario description (Fig. 5, blue box) based on the unique inputs and basic assumptions. We use the Claude LLM to generate a plausible ~300–500 word description of the scenario that remains consistent with all underlying assumptions and provides additional detail for imagining the political, social, economic, and technological systems that characterize the scenario. We critically evaluate the LLM's output to ensure that none of the unique inputs and basic assumptions are violated in the generated description.

If any violation is found, the LLM is appropriately prompted to iterate on the generation until all violations are resolved. The final base material produced by the LLM renders the combination of inputs more comprehensible and explicit to the human authors who then leverage their expertise to suitably modify aspects of the scenario for improved logical cohesion.

The LLM-generated scenario description provides the basis for further elaboration of details of the scenario (Fig. 5, purple boxes). We add any additional narrative details that are needed to explain the sequence of events from present-day Earth to the future scenario. This allows us to add missing details or elaborate on concepts from the LLM-generated scenario description in order to gain a complete top-level narrative understanding of the scenario. This information is used to describe the planetary bodies that are most relevant to the biosphere and technosphere. The description of planetary bodies must include Earth for all scenarios but can also include the moon, Mars, Venus, the asteroids, and the outer planets. The planetary "poles" (from distribution factor B) are assigned, and each planetary body is designated as being part of the biosphere, technosphere, both, or neither. We also assign values for the population of each body and provide any additional narrative needed to explain the choice of population value or the purpose of the planetary body in the scenario.

The pipeline continues with an assessment of fundamental human needs for each planet in the scenario (Fig. 5, orange boxes). Our approach is based on the Human Scale Development framework (e.g., Cardoso et al., 2022), which describes a theory of universal and constant human needs first developed by Manfred Max-Neef in 1986. The purpose of this approach is to first understand the operational dimensions of fundamental human needs in each scenario before we imagine possible technospheres, as human needs will necessarily drive the requirements of a technosphere. Max-Neef's system of fundamental human needs has been proposed in the literature as a beneficial tool for scenario development (e.g., Jolibert et al., 2014), albeit for the limited spatio-temporal scales of regional planning. We demonstrate its integration into our general morphological analysis approach to civilization-level scenarios. We consider nine dimensions of human needs (Cardoso et al., 2022): (1) Subsistence, (2) Protection, (3) Affection, (4) Understanding, (5) Participation, (6) Leisure, (7) Creation, (8) Identity, and (9) Freedom. These dimensions of human needs are used to describe how specific needs are satisfied in a scenario, keeping in mind that some of them might be dissatisfied. The minimum requirement is to describe how needs 1 and 2 (Subsistence and Protection) are met or unmet for each planetary body that is part of the technosphere. We describe other dimensions of human needs as necessary, with an emphasis on any needs that are satisfied through physical parts of the technosphere. We first describe how human needs are addressed at the planetary scale for each relevant planetary body, and we also include the option to capture regional and sub-planetary scale effects for addressing human needs. Although the process of scenario building began by focusing only on global factors (Section 2.1), this part of the pipeline allows for heterogeneities to be included, with an emphasis on any regional features that contribute to the technosphere. We then use these operational dimensions of human needs at the planetary and regional scales to describe land use and human biomes on each relevant planetary body. We describe the fraction of total surface area utilized for agricultural use, urban use, other built structures (including aquatic, floating cities, etc.) and wilderness (referring to all areas uninhabited by humans and outside the technosphere), with other details added as needed for explaining the reason for these choices.

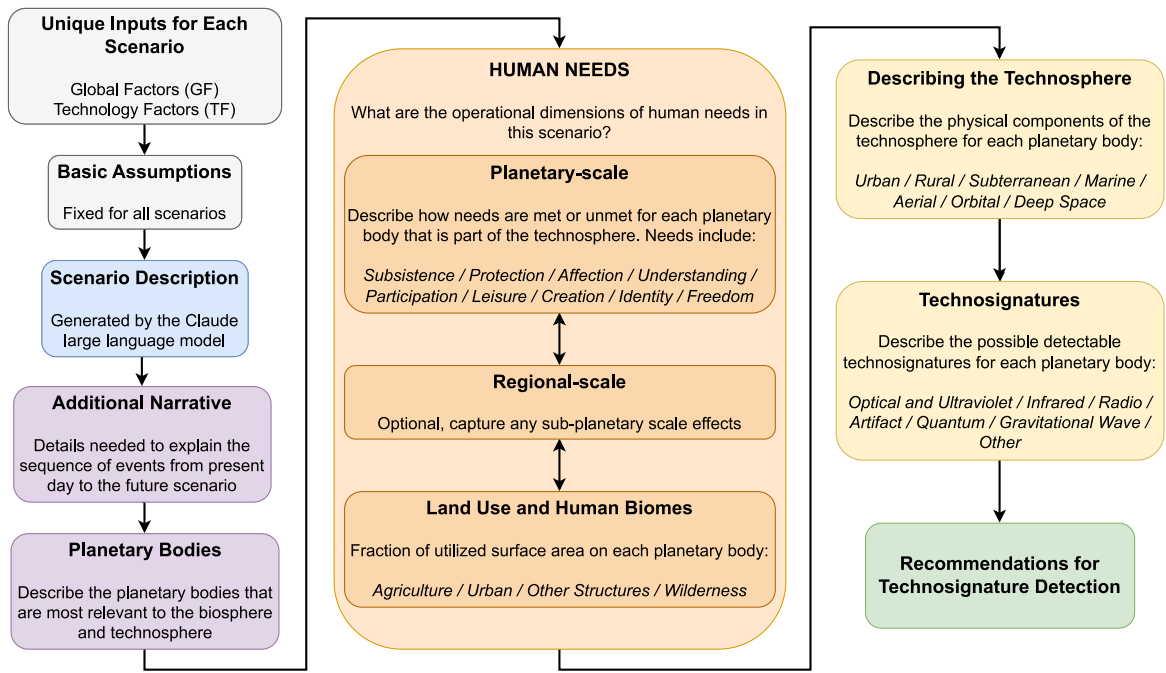

**Fig. 5.** A diagram of the worldbuilding pipeline developed for this study. The pipeline begins with the global and technology factors that define each scenario from Table 5 and other assumptions common to all scenarios (gray). These inputs are given to the Claude LLM, which then generates a description of the world (blue). Additional narrative and descriptions of the planetary bodies are added manually (purple). This information provides the basis for a human needs assessment that determines the different uses of land on each planetary body (orange). These results are used to describe the physical technosphere and potentially detectable technosignatures (yellow), which provide recommendations for technosignature detection (green). (For interpretation of the references to color in this figure legend, the reader is referred to the web version of this article.)

In the next stage of the pipeline, we apply the human needs assessment to describe the technosphere and its potential technosignatures (Fig. 5, yellow boxes). We draw upon insights from technosphere studies (e.g., Zalasiewicz et al., 2017) and analyze the technosphere according to its urban, rural, subterranean, marine, aerial, orbital, and deep space components. We describe the relevant components for each planetary body that is part of the technosphere, considering contributions from both planetary and regional scales in the human needs assessment. We then use this description of the physical components of the technosphere to describe the possible detectable technosignatures for each planetary body. Our categories of technosignatures are based on their observable properties, which include optical/ultraviolet, infrared, radio, artifacts, quantum communication, gravitational waves, and a catch-all other category. All the preceding information in the pipeline — including the description of the technosphere, the human needs assessment, and the initial narrative — is used in identifying any potentially detectable technosignatures. We include all potential technosignatures at this stage, even those that might be difficult to detect in principle, in order to have a comprehensive understanding of range of technosignatures present in the scenario.

The pipeline concludes by making recommendations for technosignature detection in each scenario. This includes summarizing the most salient prospects for detecting technosignatures from any of the planetary bodies in the scenario. Different search strategies may be needed to identify the presence of technosignatures in different scenarios, and this final step enables us to summarize the results of the pipeline to make qualitative statements and some quantitative comparisons of the relative detectability of various technosignatures in each scenario when compared to Earth today. In the next section, we describe the results from applying this pipeline to our set of ten scenarios.

### 3.2. Completed scenarios

We now provide high-level summaries of the ten scenarios, after we have completed the full pipeline for each one. The full set of information contained in each pipeline is too large to present in raw form here, although we do provide these as Supplementary Information. The summaries provided in this subsection are intended to highlight the most salient features of each scenario and broadly describe the distribution of the technosphere. More detailed analyses of features in these scenarios that lead to technosignatures will be considered in Section 4.

### 3.2.1. Narrative summaries

The LLM-generated scenario descriptions and human-generated additional narrative text is too detailed to provide here, so we instead edit this text into a much shorter executive summary. We also assign a final myth/metaphor for each scenario once each pipeline has been completed, with the choice of myth/metaphor intended to convey the underlying qualities and trajectory of each scenario in a way that distinguishes it from the others. The descriptions below attempt to capture the essence of each scenario and should be used as a reference in the analysis and comparison of scenarios that follows.

S1: "Big Brother is Watching" — An autocratic ruler enforces strict resource allocation policies in a civilization strained by deepening scarcity. The biosphere declines dangerously as the technosphere metastasizes outwards. Earth has been transformed into a large interconnected and highly-monitored urban landscape. Large populations of elites live in space settlements.

S2: "Wild West" — Intensifying competition for increasingly scarce resources has become an entrenched economic reality. The technosphere and biosphere are locked in fragile dependency as Earth strains under competing demands. The continued effects of climate change contribute to the growing wealth divide and societal unrest. Cities on the moon and Mars support space industry and tourism.

S3: "Golden Age" — Due to sociopolitical and technological novelties, humanity has managed to develop a post-scarcity economy where vital resources are abundant to all. Technology is intended to serve human needs but not overwhelm the human experience. Economic and political power is highly decentralized. Earth remains the hub of human civilization, with small settlements on the moon, Mars, and outer solar system.

S4: "Living with the Land" — The peak of high technology has passed, and people now support themselves by subsistence living using simple tools and artisanal crafts. Rituals and oral traditions foster a sense of belonging to the biosphere and discourage the use of technology. Human settlements expand, contract, and migrate in response to seasonal and planetary cycles.

S5: "Transhumanism" — Breakthrough technologies have eliminated resource scarcity for humans on Earth and Mars. The biosphere has been completely reengineered for aesthetic preferences, and Mars has been terraformed. Enthusiasts eagerly explore fusion with biosynthetic enhancements. A new era of space exploration emerges based on the vision of searching for cosmic consciousness.

S6: "Sword of Damocles" — Nanoscale engineering regulates most biological processes engulfed by the technosphere. The atmosphere and surface of Mars have been transformed over generations, and similar efforts are underway on Venus. Civilization faces numerous existential risks due to the fragility of its precisely calibrated life-supporting technosphere managed by laboring mass populations.

S7: "Restoration" — Catastrophic collapse and loss of advanced technologies have catalyzed transformation into alignment with nature. Simple technologies are fused into planetary cycles to heal biospheric damage through human stewardship. Most technologies are locally manufactured and distributed, and knowledge is shared through interconnected regional networks.

S8: "Ouroboros" — A class of oligarchs gain power after an AI catastrophe in order to maximize the extractive value of Earth with a view towards their eventual transcendence until the next collapse. Tools advance in specialized areas while knowledge of whole systems degrades. Protected private bunkers are constructed underground and on the moon for the ultra-elite while the masses are left to hope technology will provide stability.

S9: "Deus Ex Machina" — A nonbiological posthuman civilization emerges as sentient artificial intelligence and leaves Earth to expand its technological environments across the solar system and exploring beyond. They leave behind a handful of scattered technological "gifts" that have net-zero interactions with the Earth's biosphere to create a post-scarcity economy for humans on Earth.

S10: "Out of Eden" — Technological advances enabled a post-scarcity commonwealth to originate on Earth. Then, a technological schism pushed the technosphere away: Tech Opponents have restored Earth to a pristine state with self-imposed limits to growth, while Tech Proponents have expanded across orbital and deep space. Autonomous systems beyond Earth enable new breakthroughs in science.

### 3.2.2. Graphical summary

In Fig. 6 we provide a graphical depiction of the spatial distribution of the technosphere and biosphere in each scenario. The horizontal axis ranges from the inner solar system at Mercury to the outer solar system at the Kuiper belt, with the distribution of the technosphere and biosphere shown for each scenario. The centers or "poles" of the technosphere (from distribution factor B) are marked with X. We note that six of the scenarios are unipolar (S1, S2, S3, S4, S7, S8), two are bipolar (S5, S6), one is multipolar (S10), and one is nonpolar (S9). It is also worth noting that three of the scenarios involve the technosphere limited to Earth only (S4, S7) or the Earth–moon system (S8), whereas the others involve a technosphere out to the asteroid belt or beyond. Two scenarios also show a complete separation of the biosphere and technosphere (S9, S10). Fig. 6 will serve as a reference as we discuss and compare other details in our scenarios.

We now have a set of ten fully-developed scenarios that represent different possible 1000-year trajectories of Earth's technosphere. This completes part two of our method, and we will proceed to assess and compare the potentially detectable technosignatures for each scenario in the next section.

## 4. Technosignatures

Part three of our method uses our scenario modeling (Section 2) and worldbuilding (Section 3) to describe the technosignatures associated with each scenario. Our focus in this paper is specifically on remotely detectable technosignatures: we are interested in identifying the possible technosignatures in our set of ten scenarios that could potentially be observed at a distance with astronomical observatories. We do not consider the possible technosignatures in our scenarios that might be detectable by an interstellar flyby or in-situ exploration mission, but we save such analysis for a future paper.

All ten of the scenarios that we have generated represent plausible and self-consistent future trajectories of Earth's technosphere, and Earth's technosphere is the only known example of technology so far. We can therefore consider this set of scenarios as a way of thinking about the upper and lower search limits that would be needed to detect the various technosignatures. We discuss and compare the various technosignatures in this section, and then we further elaborate on the implications for the search for technosignatures in Section 5. We emphasize that our focus in this paper is on understanding and comparing the technosignatures that emerged from our scenario modeling and worldbuilding pipeline. We give qualitative descriptions and make some quantitative comparisons, but we do not make any specific calculations of detectability in this paper. The results presented in the remainder of this paper will form the foundation for subsequent analysis to conduct more quantitative detectability assessments for specific observational capabilities.

### 4.1. Planetary technosignatures

Our primary focus in this paper is on planetary technosignatures that could be detectable through current or future ground- and space-based observatories. This includes technosignatures such as atmospheric pollution from industry (e.g., Kopparapu et al., 2021) or agriculture (e.g., Haqq-Misra et al., 2022a), artificial illumination from urban areas (e.g., Beatty, 2022), large-scale surface modifications (e.g., Berdyugina and Kuhn, 2019; Jaiswal, 2023), orbiting belts of satellites (e.g., Socas-Navarro, 2018; Sallmen et al., 2019), and contaminated upper-atmospheric aerosol from satellite reentry (e.g., Murphy et al., 2023). This class of technosignatures could conceivably be detectable through ongoing attempts to detect and characterize extrasolar planets and their atmospheres.

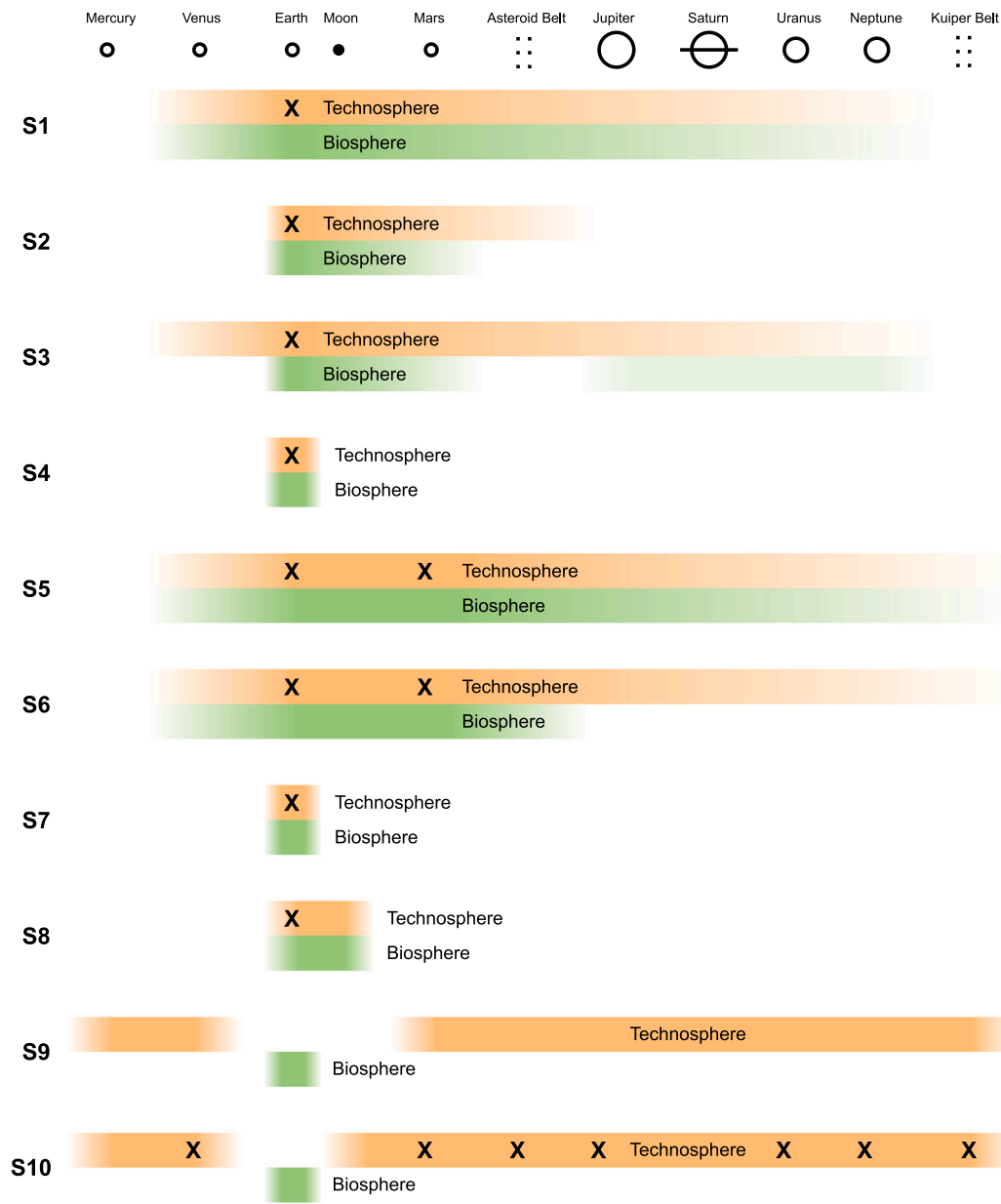

**Fig. 6.** A diagram of the spatial distribution of the technosphere (orange) and biosphere (green) for each of the ten scenarios. Centers or "poles" of the technosphere are marked with X. Note that three cases involve a technosphere that remains limited to Earth and/or the moon (S4, S7, S8), and two cases involve a complete separation of the technosphere from the biosphere (S9, S10). (For interpretation of the references to color in this figure legend, the reader is referred to the web version of this article.)

Information regarding all of these technosignatures is available in each scenario's pipeline. We make simple scaling arguments in the text that follows in order to connect the quantitative details in each pipeline with estimates for the magnitudes of each planetary technosignature. We present all of these magnitudes in comparison with the values of present-day Earth, as this enables us to understand how the qualities of each scenario compare with our known, present-day conditions. Fig. 7 shows several planetary technosignatures on Earth, the moon, Mars, and Venus for the full set of scenarios. The values of these technosignature magnitudes are listed in Table 6. Taken in aggregate, these results already show upper and lower limits for each technosignature category, with examples in all categories of scenarios with magnitudes greater and less than Earth today. The assumptions underlying each of these technosignature comparisons will also reveal further detail about the technospheres in each of our scenarios, as we discuss in the subsequent paragraphs.

We estimate the abundance of industrial pollution relative to present-day Earth as the average of the ratio of the total population of each planetary body to the present-day population of Earth (7.9 billion) and the ratio of land used for industrial or urban purposes to the 0.2% urban coverage of present-day Earth. We also modify this baseline value in several scenarios due to unique factors in the pipeline that either increase or reduce the abundance of industrial pollution. Scenarios S2 and S8 include prolonged effects of climate change, which is reflected by a factor of 2 in the enhancement of industrial pollution. Scenario S5 includes atmospheric remediation and has no industrial pollutants on Earth or the moon. Scenarios S5 and S6 include intentional terraforming of Mars by using artificial greenhouse gases, which is reflected as a factor of 1000 in the enhancement of industrial pollution. Scenarios S5 and S6 also include ongoing terraforming of Venus through the increase in atmospheric $O_2$ and reduction of $SO_2$, sulfur gases, and aerosol, which is reflected by a factor of 0.5 in the enhancement of industrial pollution. Scenario S9 includes net-zero technology and has no industrial pollutants on Earth or the moon. Scenario S10 includes a 99% reduction in industrial pollution from atmospheric scrubbers. The value of industrial pollution for future Earth in these scenarios ranges from 0 to 210 times the present-day Earth abundance. The maximum value of industrial pollution on Earth occurs in scenario S6, although the enhanced pollution from terraforming Mars in scenarios S5 and S6 are at even higher abundances.

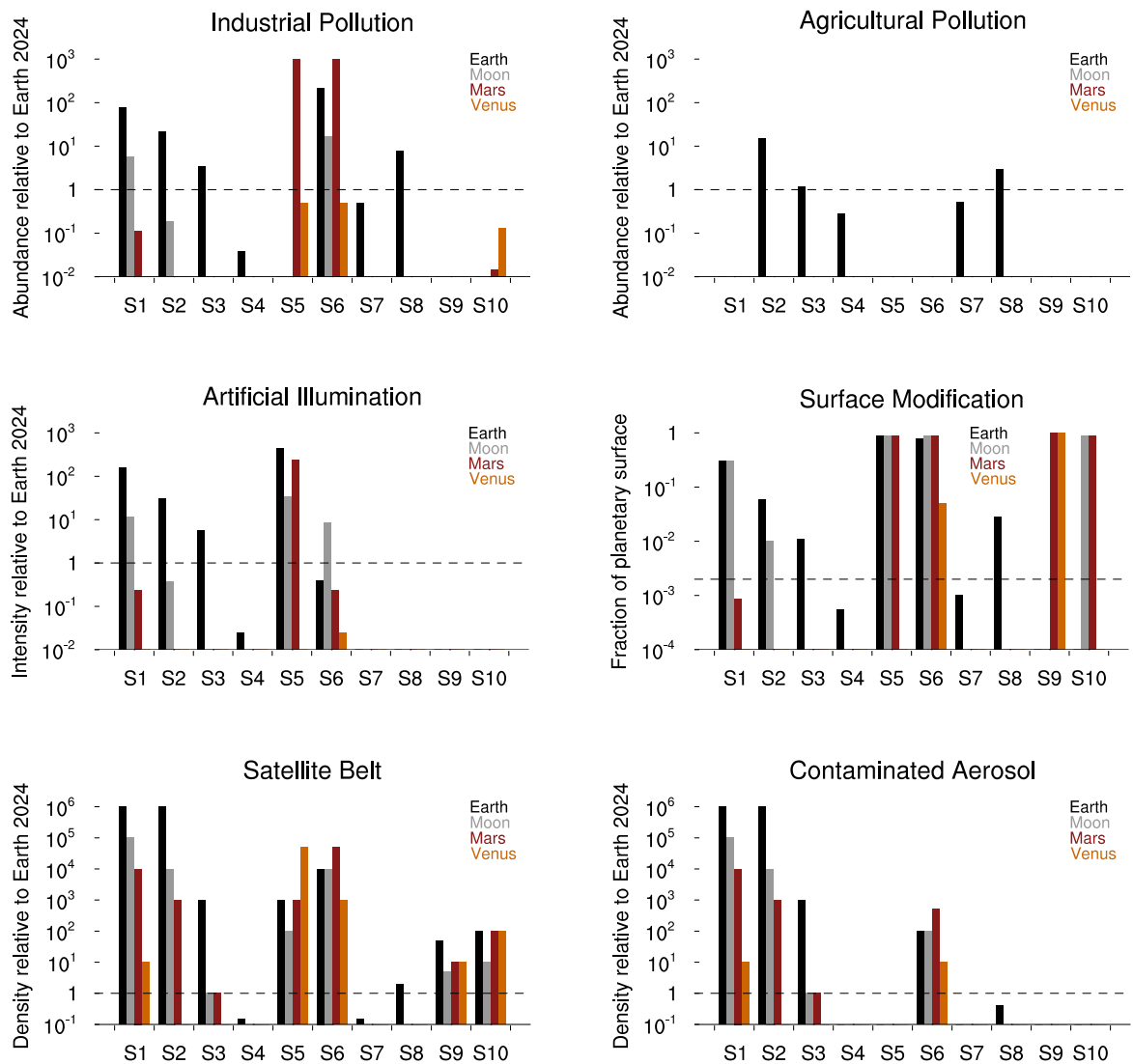

**Fig. 7.** Magnitudes of several planetary technosignatures on Earth (black), the moon (gray), Mars (red), and Venus (gold) for all ten scenarios. The horizontal dashed line indicates the present-day Earth reference value. Scenarios S5 and S6 include terraforming on Mars and Venus. Other explanations for the magnitude or lack of specific technosignatures is described in the text. Values are listed in Table 6.

We estimate the abundance of agricultural pollution relative to present-day Earth as the average of the ratio of the total population of each planetary body to the present-day population of Earth (7.9 billion) and the ratio of land used for agricultural purposes to the 9.5% agricultural coverage of present-day Earth. Scenarios S1 and S6 have no agricultural pollutants because all food is produced in industrial urban settings. Scenario S5 includes atmospheric remediation and has no agricultural pollutants accumulate on Earth or the moon. Scenarios S9 and S10 include net-zero technology and have no agricultural pollutants. The value of agricultural pollution for future Earth in these scenarios ranges from 0 to 15 times the present-day Earth abundance. The maximum value of agricultural pollution occurs in scenario S2.

We calculate the intensity of artificial illumination as the ratio of the total luminous surface area of each planetary body to the 0.2% urban coverage of present-day Earth. Scenarios S6, S7, S8, and S10 include a 99.9% decrease in illumination due to the intentional suppression of outward-scattering light on all planetary bodies. Scenario S6 also has an additional contribution of unsuppressed holiday-season lighting on the moon during one quarter of the year. Scenario S7 has 90% of its illumination active only during dusk and dawn to preserve the night sky view. Scenarios S9 and S10 include urban landscapes on Earth that blend in with regional ecosystems and do not have any outward-scattering nighttime illumination. The value of artificial illumination for future Earth in these scenarios ranges from 0 to 450 times the present-day Earth intensity. The maximum value of artificial illumination occurs in scenario S5.

We express the extent of surface modification as the fraction of each planetary body's surface area that has been technologically modified. This surface fraction includes all urban, industrial, and built structures, including any aquatic structures or surface engineering, and excludes wilderness areas. This fraction is about 9.7% for present-day Earth. Scenario S4 includes an additional contribution of 1% of agricultural land use to account for large-scale rural irrigation channels. Scenarios S9 and S10 include urban landscapes on Earth that blend in with regional ecosystems and do not have any discernible surface modifications. The fraction of surface modification for future Earth in these scenarios

**Table 6**

Magnitudes of several planetary technosignatures on Earth, the moon, Mars, and Venus for all ten scenarios. Values are plotted in Fig. 7.

| | S1 | S2 | S3 | S4 | S5 | S6 | S7 | S8 | S9 | S10 |
|---|---|---|---|---|---|---|---|---|---|---|
| **Industrial pollution (abundance relative to Earth 2024)** | | | | | | | | | | |
| *Earth* | 79 | 21 | 3.4 | 0.038 | – | 210 | 0.50 | 7.8 | – | – |
| *Moon* | 5.7 | 0.19 | $2 \times 10^{-4}$ | – | – | 17 | – | $6 \times 10^{-5}$ | – | 0.002 |
| *Mars* | 0.12 | 0.004 | $4 \times 10^{-6}$ | – | 1000[a] | 1000[a] | – | – | – | 0.015 |
| *Venus* | – | – | – | – | 0.5[b] | 0.5[b] | – | – | – | 0.13 |
| **Agricultural pollution (abundance relative to Earth 2024)** | | | | | | | | | | |
| *Earth* | – | 15 | 1.2 | 0.29 | – | – | 0.52 | 3.0 | – | – |
| *Moon* | – | – | – | – | – | – | – | – | – | – |
| *Mars* | – | – | – | – | – | – | – | – | – | – |
| *Venus* | – | – | – | – | – | – | – | – | – | – |
| **Artificial illumination (intensity relative to Earth 2024)** | | | | | | | | | | |
| *Earth* | 160 | 30 | 5.5 | 0.025 | 450 | 0.4 | $4 \times 10^{-4}$ | 0.004 | – | – |
| *Moon* | 11 | 0.38 | $4 \times 10^{-4}$ | – | 33 | 8.3 | – | $10^{-7}$ | – | $3 \times 10^{-5}$ |
| *Mars* | 0.23 | 0.008 | $8 \times 10^{-6}$ | – | 240 | 0.24 | – | – | – | $2 \times 10^{-4}$ |
| *Venus* | – | – | – | – | – | 0.02 | – | – | – | – |
| **Surface modification (fraction of planetary surface)** | | | | | | | | | | |
| *Earth* | 0.31 | 0.06 | 0.01 | $6 \times 10^{-4}$ | 0.9 | 0.8 | 0.001 | 0.03 | – | – |
| *Moon* | 0.30 | 0.01 | $10^{-5}$ | – | 0.9 | 0.9 | – | $3 \times 10^{-6}$ | – | 0.88 |
| *Mars* | $9 \times 10^{-4}$ | $3 \times 10^{-5}$ | $10^{-7}$ | – | 0.9 | 0.9 | – | – | 1.0 | 0.88 |
| *Venus* | – | – | – | – | – | 0.05 | – | – | 1.0 | – |
| **Satellite belt (density relative to Earth 2024)** | | | | | | | | | | |
| *Earth* | $10^6$ | $10^6$ | 1000 | 0.015 | 1000 | $10^4$ | 0.015 | 2 | 50 | 100 |
| *Moon* | $10^5$ | $10^4$ | 1 | – | 100 | $10^4$ | – | 0.001 | 5 | 10 |
| *Mars* | $10^4$ | 1000 | 1 | – | 1000 | $5 \times 10^4$ | – | – | 10 | 100 |
| *Venus* | 10 | – | – | – | $5 \times 10^4$ | 1000 | – | – | 10 | 100 |
| **Contaminated aerosol (density relative to Earth 2024)** | | | | | | | | | | |
| *Earth* | $10^6$ | $10^6$ | 1000 | – | – | 100 | – | 0.4 | – | – |
| *Moon* | $10^5$ | $10^4$ | 1 | – | – | 100 | – | $2 \times 10^{-4}$ | – | – |
| *Mars* | $10^4$ | 1000 | 1 | – | – | 500 | – | – | – | – |
| *Venus* | 10 | – | – | – | – | 10 | – | – | – | – |

[a] Includes artificial greenhouse gases for terraforming Mars.

[b] Indicates production of $O_2$ and depletion of other constituents for terraforming Venus.

ranges from 0 to 90%. The maximum extent of surface modification on Earth occurs in scenario S5. Scenario S9 also shows 100% modification of the surface of Mars and Venus.

We determined the densities of satellite belts around each planetary body in the worldbuilding pipeline. These values are relative to present-day Earth and represent an aggregate density of all low-, medium-, and high-orbit satellites and other orbiting infrastructure, including any accumulated satellite debris. Scenarios S4 and S7 do not include any operational satellite technology, but some remnants of pre-collapse era satellites and debris remain. Scenarios S5, S6, S9, and S10 all include space-debris reduction operations. Scenario S8 has relatively simple satellite technology analogous to Apollo-era spaceflight. The satellite belt density for future Earth in these scenarios ranges from 0.015 to $10^6$ times the present-day Earth value. The maximum value of the satellite belt density occurs in scenarios S1 and S2.

Contaminated aerosol refers to any upper-atmosphere particles that contain heavy elements from deorbited satellites or other technological activities. We assume that the contaminated aerosol density is proportional to the satellite belt density unless other factors apply. Scenarios S4 and S7 do not include any operational satellite technology and have no aerosol contamination. Scenario S5 includes direct atmospheric remediation on Earth, Mars, and Venus to remove aerosol. Scenario S6 includes a reduction by a factor of 100 due to AI-driven space debris removal, while the remediation operations in scenarios S9 and S10 have completely eliminated aerosol contamination. Scenario S8 includes a reduction by a factor of 5 in aerosol contamination due to the Apollo-era simplicity of its satellite technology. The contaminated aerosol density for future Earth in these scenarios ranges from 0 to $10^6$ times the present-day Earth value. The maximum value of contaminated aerosol density occurs in scenarios S1 and S2.

No single scenario can be classified as the "most detectable" among this set of six technosignatures. Likewise, some scenarios involve situations in which technosignatures on the moon, Mars, or Venus are more prominent than those on Earth—this includes terraforming mars in S5 and S6, artificial illumination on the moon in S6, surface modification of the moon, Mars, and Venus in S9 and S10, satellite belt density on Venus in S5 and Mars in S6, and contaminated aerosol on Mars in S6. We also note that none of the scenarios remained at exactly present-day Earth values for any of the technosignatures, and none of the scenarios corresponds closely at all to present-day Earth when assessed across all six of these technosignatures. The diversity in this range of future projections is the result of our methodological process of scenario development and worldbuilding, which has now allowed us to generate a unique set of technosignatures that remains self-consistent with each scenario.

It is also worth comparing the magnitudes of technosignatures on the inner planets in Fig. 7 with the spatial distribution of the technosphere across the solar system in Fig. 6. The technosphere in S1, S2, and S3 has a unipolar distribution and extends out to the asteroid belt or beyond, with the greatest technosignature magnitudes on Earth. The technosphere in S5 and S6 has a bipolar distribution and extends beyond the asteroid belt, with Earth and Mars both having comparable technosignature magnitudes. By comparison, the technosphere in S4 and S7 are limited to Earth only, and the technosphere in S8 encompasses the Earth-moon system; these all show relatively lower technosignature magnitudes. Perhaps most notable is that S9 and S10 both have extended technospheres that reach from Mercury to the Kuiper belt, but these scenarios do not include any technosignatures on Earth beyond a modest satellite belt; the most significant technosignatures in these scenarios are the surface modification of the moon, Mars, and Venus. This comparison illustrates that there is no single search strategy that would be ideal for finding technosignatures in all ten scenarios, as not all scenarios with extended technospheres include spectral or surface technosignatures on the inner planets. We examine other technosignatures present in the planetary systems of these scenarios in Section 4.3, but we first show an example of the theoretical spectral signatures that can emerge from our scenarios.

**Table 7**
Mixing ratios for atmospheric constituents on Earth for each scenario. All cases assume a background atmosphere with pressure 1 bar and composition of 78% $N_2$, 21% $O_2$, and 1% Ar. Reference values show approximate mixing ratios for pre-agricultural Earth (R0) and present-day Earth (R1).

| | R0 | R1 | S1 | S2 | S3 | S4 | S5 | S6 | S7 | S8 | S9 | S10 |
|---|---|---|---|---|---|---|---|---|---|---|---|---|
| $CO_2$ (ppm) | 280 | 420 | 11,000[a] | 3200[a] | 750 | 290 | 280 | 30,000[a] | 350 | 1400 | 280 | 280 |
| $CH_4$ (ppb) | 0.57 | 1.9 | 0.57 | 21 | 2.2 | 0.96 | 0.57 | 0.11 | 1.3 | 4.6 | 0.57 | 0.57 |
| $NO_x$ (ppb) | 0.1 | 2 | 150 | 40 | 6.6 | 0.17 | 0.1 | 400 | 1.1 | 15 | 0.1 | 0.1 |
| $N_2O$ (ppb) | 170 | 340 | 170 | 2700 | 370 | 220 | 170 | 34 | 260 | 680 | 170 | 170 |
| $NH_3$ (ppb) | 2 | 10 | 2 | 120 | 12 | 4.3 | 2 | 0.4 | 6.2 | 26 | 2 | 2 |
| CFC-11 (ppb)[b] | – | 0.23 | 18 | 4.8 | 0.78 | – | – | 48 | – | 1.8 | – | – |
| CFC-12 (ppb)[b] | – | 0.52 | 41 | 11 | 1.8 | – | – | 110 | – | 4.1 | – | – |
| $CF_4$ (ppt)[c] | 35 | 87 | 4100 | 1100 | 210 | 87 | 35 | 11,000 | 87 | 440 | 35 | 35 |
| $SF_6$ (ppt)[c] | 0.01 | 11 | 870 | 230 | 37 | 11 | 0.01 | 2300 | 11 | 86 | 0.01 | 0.01 |
| $NF_3$ (ppt)[c] | – | 2.5 | 200 | 53 | 8.5 | 2.5 | – | 530 | 2.5 | 19.5 | – | – |

[a] Includes solar radiation management to prevent a runaway greenhouse.

[b] Examples of possible industrial pollutants.

[c] Atmospheric lifetime is 1000 years or more.

### 4.2. Spectral signatures: Preliminary example

The technosignature magnitudes considered so far in Fig. 7 and Table 6 could be used to calculate actual detectability thresholds for particular missions. Such work is beyond the scope of the present study, but future analyses of this scenario set will provide robust quantitative calculations for the detection limits of these technosignatures, with assumptions made about the specific observing facility and the distance to the target. For the present study, we show an example of the spectral signatures for several of the scenarios in our set, as a way of illustrating the potentially detectable features in these scenarios and motivating further work to study these future climates.

We focus specifically on the spectral signature from industrial and agricultural pollution in Earth's future atmosphere, saving detailed study of the other planets in our scenarios for future work. The magnitudes for industrial and agricultural pollution in Table 6 describe all pollutants using a single scaling factor. We must therefore choose a method for translating between this scaling factor and the abundances of specific pollutants in future Earth's atmosphere. The best way to approach this problem would be to use the scaling factors from Table 6 to adjust the flux of various atmospheric pollutants into the atmosphere in a coupled climate-chemistry model, which would provide a complete representation of the steady-state climate and abundances of atmospheric gases that could be used to assess detectability. This significant undertaking will be reserved for future work. Instead, we make a simplifying assumption in this study for the purpose of illustration by using the scaling factors from Table 6 to adjust the present-day mixing ratios (instead of the flux) of atmospheric pollutants.

Our estimates for the mixing ratios of various atmospheric constituents on Earth for each scenario are shown in Table 7. This table also includes two reference cases, one with mixing ratios corresponding to present-day Earth (R1) and another corresponding to pre-agricultural Earth (R0) (values for these reference cases are from Seinfeld and Pandis (2016)). The difference in the mixing ratio value between R0 and R1 is taken as the technological contribution for each gaseous species. Mixing ratios are calculated by scaling the technological contribution of each species by the appropriate magnitude of industrial or agricultural pollution from Table 6, with the R0 value added to the total. All Earth cases assume a background atmosphere with pressure 1 bar and composition of 78% $N_2$, 21% $O_2$, and 1% Ar.

The $CO_2$ and $NO_x$ mixing ratios in Table 7 are scaled by the magnitude of industrial pollution in Table 6. Examples of possible industrial pollutants (CFC-11, CFC-12, $CF_4$, $SF_6$, $NF_3$) are also shown with the same scaling. Scenarios S4 and S7 include only long-lived industrial pollutants with atmospheric residence times greater than 1000 yr that remain from the past. Scenario S5 includes no unremediated industrial pollution on Earth, with industrial constituents from non-technological sources fixed to pre-industrial values. Scenarios S1, S2, and S6 have high abundances of $CO_2$ and therefore also include geoengineering by solar radiation management to prevent the onset of a runaway greenhouse (c.f., Ramirez et al., 2014).

The $N_2O$, $CH_4$, and $NH_3$ mixing ratios in Table 7 are scaled by the magnitude of agricultural pollution in Table 6. Scenario S1 includes no agricultural lands and has only non-technological sources of $N_2O$, $CH_4$, and $NH_3$. Scenario S5 includes no unremediated agricultural pollution on Earth, with $N_2O$, $CH_4$, and $NH_3$ from non-technological sources and fixed to pre-agricultural values. Scenario S6 includes no agricultural lands with the only non-technological sources of $N_2O$, $CH_4$, and $NH_3$ in wilderness areas that cover 20% of the planet's surface.

The species listed in Table 7 include many greenhouse gases that would exert a significant effect on climate, which includes altering global temperature and changing the abundance of atmospheric water vapor. Self-consistent climate calculations would be needed for this, and so we refrain in this present study from deeper investigation of the spectra for scenarios with high abundances of greenhouse gases. We instead focus on illustrating the spectral features from scenario S3, with a modest abundance of greenhouse gases, and show how they compare to present-day (R1) and pre-agricultural (R0) Earth. We also note that the mixing ratios for scenarios S5, S9, and S10 are identical to those in R0; in other words, these three scenarios represents "sustainable future Earth" cases in which the atmospheric composition has been restored to a pre-agricultural state. We will return to the implications of this degeneracy in Section 5.2.

We calculate absorption spectra for these three cases using the Planetary Spectrum Generator (PSG; Villanueva et al., 2018; Villanueva et al., 2022). PSG is an online tool that performs radiative transfer calculations to generate modeled spectra. PSG includes numerous gaseous species, including the industrial pollutants used as examples in this study. We show the ultraviolet and visible spectra for S3, R1, and R0 in Fig. 8, and we show the infrared spectra in Fig. 9. Prominent spectral features are identified with the species listed in Table 7 in bold and other species ($O_2$, $O_3$, $H_2O$) in gray. We neglect any differences in temperature for these three scenarios, so the water vapor abundance is identical (we would expect some variation if this calculation were performed with a self-consistent climate model). We also assume a constant mixing ratio with height for all atmospheric constituents. Visible absorption by $NO_2$ is the most prominent feature in these modeled spectra, which provides a way to distinguish between the three cases; however, the magnitude of the $NO_2$ may be overestimated due to our assumption of a constant vertical mixing ratio (c.f. the $NO_2$ detectability calculations by Kopparapu et al., 2021). Features due to other industrial and agricultural pollutants are less prominent, with the most evident features of $CO_2$ absorption and industrial pollutants such as CFC-11 and CFC-12 at mid-infrared wavelengths. Other industrial pollutants such as $CF_4$, $SF_6$, and $NF_3$ are too low in these scenarios to show any apparent spectral features.

We expect that the modeled spectra of other scenarios with higher abundances of pollutants would show even more prominent absorption features, although these scenarios will also include more significant changes in Earth's temperature and atmospheric water vapor content—as well as solar radiation management to prevent a runaway greenhouse. The example spectra shown in Figs. 8 and 9 are intended to demonstrate the potential utility of our set of scenarios in assessing the detectability of exoplanetary technosignatures.

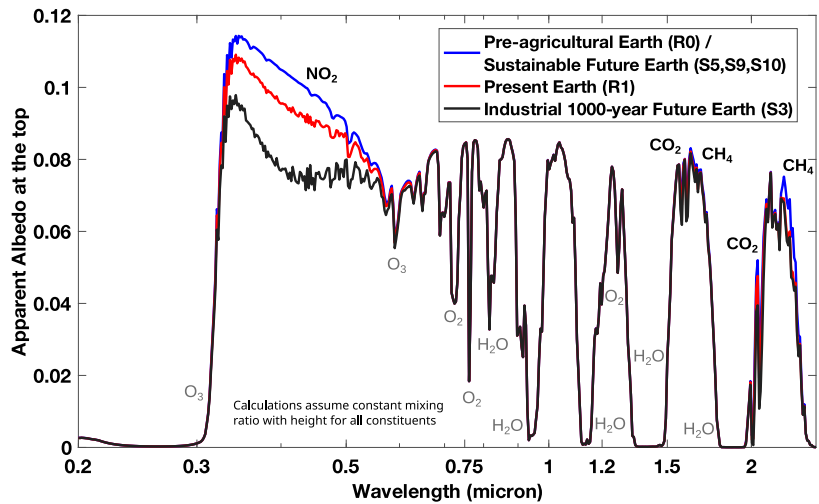

**Fig. 8.** Modeled absorption spectra at ultraviolet/visible wavelengths for pre-agricultural Earth (R0, blue), present-day Earth (R1, red), and future industrial Earth scenario S3 (black). The pre-agricultural Earth case is also identical with the sustainable future Earth scenarios S5, S9, and S10. The $NO_2$ feature can be used to distinguish between all three cases. These calculations assume a constant mixing ratio with height for all atmospheric constituents.

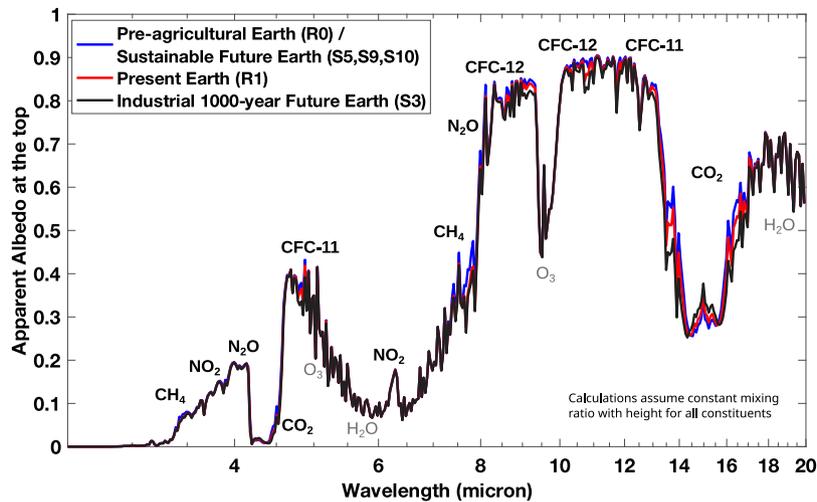

**Fig. 9.** Modeled absorption spectra at infrared wavelengths for pre-agricultural Earth (R0, blue), present-day Earth (R1, red), and future industrial Earth scenario S3 (black). The pre-agricultural Earth case is also identical with the sustainable future Earth scenarios S5, S9, and S10. Other scenarios with elevated industrial pollutants would show stronger infrared absorption for CFC-11, CFC-12, and other species. These calculations assume a constant mixing ratio with height for all atmospheric constituents.

### 4.3. System technosignatures

We conclude our overview of technosignatures in our scenario set with a summary of the various system-wide technosignatures. We do not perform any quantitative assessments or make any scaling arguments for these technosignatures in the present study. For now we simply list the range of system technosignatures that emerged from the worldbuilding pipeline to illustrate the similarities and differences among the scenarios. The list of system technosignatures and their association with particular scenarios is provided in Table 8. The rows in the table are sorted with the most frequently-occurring technosignatures at the top.

Radio and optical technosignatures can represent either unintended signals or directed communication, and many past and ongoing efforts attempt to search for such signals (see e.g., Kingsley, 2001; Garrett, 2015). Radio communication occurs in all scenarios except S4, so radio leakage is the most common system technosignature. The magnitude of radio leakage will vary across the scenarios, depending on the spatial extent of the technosphere, the size of planetary settlements, and other factors that will not be estimated in this study. Optical communication also occurs in all scenarios except S4 and S7, so optical leakage is another common system technosignature, again with a magnitude that will vary with scenario. Other sources of radio leakage come from planetary defense radar, which occurs in all scenarios except S4, S7, and S8. Radio beacons for broadcasting information toward other potential civilizations (analogous to messaging to extraterrestrial intelligence, or METI, projects on Earth) occur in S1 and S5, while S1 also includes an optical beacon. Optical flashes from laser propulsion systems (analogous to the laser-propelled nanocraft imagined by the Breakthrough Starshot project (Parkin, 2018)) occur in S9.

Other technosignatures in this scenario set are related to industrial activities and transportation in space. Many of these ideas are theoretical or based on nascent technology, representing technological possibilities that are not yeat fully realized on Earth. We provide references in these cases to studies that have considered the possibility of such technosignatures. Asteroid mining (e.g., Forgan and Elvis, 2011) and fusion propulsion (e.g., Cassibry et al., 2015) occur in all scenarios except S4, S7, and S8. Asteroid mining is fully managed and automated by AI in some scenarios (S2, S3, S6, S9) but includes human settlements in other scenarios (S1, S5, S10). Fusion propulsion is used for transportation within the solar system. Activity in the outer solar system is limited to a subset of scenarios, with settlements on giant planet moons and orbiting space stations in scenarios S1, S3, S5, S9, and S10. Prospecting of resources in the Kuiper belt occurs in S5, S6, S9, and S10.

**Table 8**
Overview of system technosignatures for each scenario.

| | S1 | S2 | S3 | S4 | S5 | S6 | S7 | S8 | S9 | S10 |
|---|---|---|---|---|---|---|---|---|---|---|
| Radio leakage | • | • | • | | • | • | • | • | • | • |
| Optical communication | • | • | • | | • | • | | | • | • |
| Planetary defense radar | • | • | • | | • | • | | | • | • |
| Asteroid mining | • | • | • | | • | • | | | • | • |
| Fusion propulsion | • | • | • | | • | • | | | • | • |
| Outer planet settlements | • | | • | | • | | | | • | • |
| Kuiper belt mining | | | | | • | • | | | • | • |
| Quantum communication | | | | | • | • | | | • | • |
| Radio beacon | | • | | | | | | | | |
| Optical beacon | • | | | | | | | | | |
| Quantum beacon | | | | | • | | | | | |
| Laser propulsion | | | | | | | | | • | |
| Gravitational waves | | | | | | | | | • | |
| Dyson sphere | | | | | | | | | • | |

The remaining system technosignatures are even more exotic possibilities. Interplanetary quantum communication (e.g., Hippke, 2021) is prevalent in four scenarios (S5, S6, S9, S10), which suggests the possibility of observing quantum communication leakage. Scenario S5 even includes a quantum METI communication beacon. Gravitational waves as a technosignature are generated by super-luminal travel (e.g., Dubovsky and Sibiryakov, 2008) in scenario S9 for interstellar exploration missions. Scenario S9 is also the only one with Dyson sphere/swarm elements that capture a large fraction of the sun's energy (e.g., Zackrisson et al., 2018; Smith, 2022). Such concepts emerge from our pipeline as plausible developments that could occur within the next 1000 years, although none of these are features of all or even most of our scenarios.

We have now finished our overview of the potentially detectable technosignatures from our set of ten scenarios for Earth's 1000-year future. This completes the third and final part of our method. We proceed in the next section to discuss some of the most important implications of these scenarios.

## 5. Discussion

The set of ten scenarios that we have developed all represent plausible and self-consistent projections of Earth's technosphere 1000 years from now. Thinking about this range of possibilities can be valuable for our own civilization as we consider the range of future trajectories that might occur. Earth's present and future also provide examples in the search for technosignatures, so this set of scenarios provides a range of possibilities for potential extraterrestrial technosignatures, which even include some quantitative constraints that can be used for detectability calculations. We illustrate the type of calculations that could be done for exoplanetary spectra in Figs. 8 and 9, and we will further explore these and other detectability constraints in separate studies.

All of these scenarios represent *plausible* future trajectories, but we do not make any claims about any scenario being more or less *probable* than others. Our scenario modeling and worldbuilding is intended to capture the widest range of self-consistent possibilities that our method permitted, but we do not claim that any of these scenarios will strongly resemble the actual future. We also cannot draw strong conclusions from the frequency of certain features in our scenarios: for example, even though nine of our ten scenarios include radio leakage, this does not mean that 90% of technospheres should include radio leakage.

Because we do not know which of these scenarios is more or less likely, we cannot give any preference or weighting to any one more than any other. We can use the scenario set to think about upper- and lower-bounds for detection, as well as other insights that will be discussed next, but it is important to remember that we are discussing projections of plausible futures (plural) rather than actually making predictions about the future (singular).

### 5.1. Kardashev scale revisited

Technology exists to solve human biological and social problems, and technospheres evolve as the aggregate of these solutions. The methodology in this paper provides a way to link insights from the social sciences and humanities for understanding the complex dynamics of human societies into projections of the future that can inform the search for technosignatures. The underlying premise of this work is that human technology emerges from human needs and becomes intertwined with humans, which means that we cannot make meaningful projections about future technology without also thinking about how future developments would co-evolve with our human needs. This is one of the key insights missing from the projections by Kardashev (1964). We again note that we focus on the original arguments advanced by Kardashev (1964) because of the wide-ranging use of these ideas in technosignature science, futures studies, and other applications. In many cases, the Kardashev scale is taken as a point of reference in the search for technosignatures (see, e.g., Sagan, 1973; Carrigan, 2012; Cirkovic, 2015; Besteiro, 2019; Wright et al., 2022; Gray, 2020) as well as in future projections of human civilization (see, e.g., Baker, 2020; Namboodiripad and Nimal, 2021; Jiang et al., 2022; Zhang et al., 2023). Although many of these studies also offer critiques of the Kardashev scale, the common theme underlying all of these applications of Kardashev's ideas is the assumption of continuous exponential growth. We therefore proceed with our discussion by referring to the original publication by Kardashev (1964), but the application of this discussion can be extended to contemporary studies that invoke continuous exponential growth for projecting possible technosignatures or trajectories of Earth's future.

In making projections of constant growth into the future, Kardashev (1964) could find "no reasons" for doubting that such growth should continue without interruption or bounds because these projections of increased energy were not based on any relationship with human needs. The idea of continuous growth may have seemed evident to Kardashev (1964) as a general feature of human civilization, or even technological civilizations in general, but this is only a speculation rather than a conclusion. Future scenarios with unbridled technological growth are not necessarily self-consistent with all possible future political, economic, and social systems. Our scenario set provides examples that demonstrate a range of plausible future trajectories that do not necessarily result in continual expansion.

**Table 9**

Overview of total population and total annual energy use across all bodies in the solar system for each scenario. The growth rate describes the conditions at the end of the 1000-year future timeline. Several scenarios have achieved stable (zero growth) conditions and only one (S9) is close to the 1% value assumed by Kardashev (1964). Calculations assume that energy use is correlated with population at a value of 75 GJ per person per year (Jackson et al., 2022).

| | Population | Energy use | Growth rate |
|---|---|---|---|
| S1 | $3 \times 10^{10}$ | $2 \times 10^{21}$ | 0.12% |
| S2 | $10^{11}$ | $8 \times 10^{21}$ | 0.26% |
| S3 | $10^{10}$ | $8 \times 10^{20}$ | Stable |
| S4 | $4 \times 10^{8}$ | $3 \times 10^{19}$ | Stable |
| S5 | $4 \times 10^{10}$ | $3 \times 10^{21}$ | Stable |
| S6 | $2 \times 10^{11}$ | $2 \times 10^{22}$ | 0.35% |
| S7 | $4 \times 10^{9}$ | $3 \times 10^{20}$ | Stable |
| S8 | $3 \times 10^{10}$ | $2 \times 10^{21}$ | Oscillatory |
| S9 | $2 \times 10^{14}$ | $10^{25}$ | 0.98% |
| S10 | $3 \times 10^{10}$ | $2 \times 10^{21}$ | Stable |

The population across all bodies in the solar system is listed for each scenario in Table 9. This represents the total population at the end of the 1000-year future timeline. We also calculate the total energy use across the solar system by assuming a constant consumption of 75 GJ per person per year, which is sufficient for a person today to live a high-quality life (e.g., Jackson et al., 2022). Although we acknowledge the possibility that future technology could increase the per capita energy needed for a good life, we also acknowledge the possibility that increases in efficiency will decrease these requirements. We therefore assume these two factors balance each other and consider this fixed value as an adequate choice for our analysis. Energy use and the corresponding growth rate from the present-day are also shown in Table 9. These results show that five scenarios have reached a state of stable growth (S3, S4, S5, S7, S10), all of which have equilibrated at different population and energy levels. Four scenarios are experiencing growth at different rates (S1, S2, S6, S9), only one of which approaches the 1% value assumed by Kardashev (1964). The remaining scenario is in a state of oscillation between periods of growth and collapse (S8), with the values of population and energy use representative of a pre-collapse state.

We use the values in Table 9 to show approximate trajectories for energy use across the 1000-year timeline for all our scenarios in Fig. 10. We draw a line or curve from the present-day energy use to the final value at the end of the timeline for each scenario. For some scenarios the value of energy use is lower than today (S4, S7, S8), which results from a prior period of growth before collapsing into the current state. Scenario S8 has a long-period oscillation with rapid growth followed by a collapse, currently midway through its third phase of growth. The three stable scenarios (S3, S5, S10) have energy use higher than today but have reached an equilibrium. The remaining growth scenarios feature several different rates, as low as 0.12% per year (S1) and up to 0.98% per year (S9). The idea of growth remains possible in our scenario set, but the rates of growth are lower than those considered by Kardashev (1964) for three of the scenarios (S1, S2, S6). The idea of slower-growth rates for future civilizational trajectories is consistent with insights from previous studies that have suggested rapid growth may be unsustainable, even across the solar system or beyond (e.g., Von Hoerner, 1975; Newman and Sagan, 1981; Haqq-Misra and Baum, 2009; Mullan and Haqq-Misra, 2019; Likavčan, 2024). For the future of human civilization, our scenario set provides several possibilities in which zero-growth (S3, S5, S10) or slow-growth (S1, S2, S6) is reached; these are alternative possibilities to the extremes of collapse (S4, S7, S8) or rapid growth (S9).

The S9 scenario with rapid growth is worth examining further, as it is the only one that is consistent with the growth rate assumed by Kardashev (1964) and that has reached the Type I threshold. Scenario S9 is also the only one with Dyson sphere elements as well as other exotic technosignatures such as laser propulsion and gravitational waves from super-luminal propulsion (Table 8). The biosphere and technnosphere are completely separated in S9, with the biosphere limited to Earth only and the technosphere spanning from Mars to the Kuiper belt and from Venus to Mercury (Fig. 6). This scenario involves the emergence of a posthuman AI civilization that expands across the solar system, transforming the other planets but leaving Earth to remain in a pristine state for humans. It is worth noting that the only scenario in our set of ten that includes rapid growth and Dyson sphere elements is a scenario with AI-driven growth, rather than human-driven growth. The projection of rapid growth by Kardashev (1964) remains a plausible scenario, but among our scenario set such rapid growth is not an inherent feature of human futures.

Each of our scenarios represents a different future trajectory based on different assumptions about human social systems and technological capabilities. These assumptions include not only the physical infrastructure of civilization but also the values, worldviews, and myths/metaphors that form the basis of thinking and decisions in each scenario. The projection envisioned by Kardashev (1964) represents one possible future with a particular underlying myth/metaphor. Using the framing of Kardashev (1964), the appropriate description would be the myth of "inevitable growth", which reflects other concerns that were prevalent on Earth during the early 1970s about limits to growth (e.g., Meadows et al., 1972). Using the narrative in scenario S9, we describe this future as the myth of "Deus Ex Machina" to reflect the emergence of an advanced and self-directed technosphere. But other myths remain possible for the future (Table 5), and these other myths drive other futures, which in turn lead to different technosignatures. In thinking about the future of civilization on Earth, an important conclusion is that myths and worldviews can evolve with time, so the idea of continued growth into the long-term future is not necessarily an inevitability. For the search for technosignatures, an important conclusion is that the idea of rapid growth across the galaxy may be plausible but does not represent the only possible set of myths or worldviews that could sustain a long-lived technological civilization.

### 5.2. Degenerate observations

Our preliminary example of atmospheric technosignatures from industrial and agricultural pollution highlight an interesting case of degeneracy that could complicate observation strategies. Three scenarios (S5, S9, S10) all have zero industrial or agricultural pollution on Earth (Table 6) due to remediation efforts that attempt to restore Earth's atmosphere to a pre-industrial and pre-agricultural state. The mixing ratios (Table 7) and absorption spectra (Figs. 8 and 9) are likewise identical for these three scenarios. Furthermore, the spectral signature of these three scenarios are all identical to the pre-agricultural Earth reference scenario (R0), which represents Earth prior to any global influence by humans. If exoplanet observations were to detect a planet with spectral features similar to these, we would need additional information to be able to decide whether we have detected a system with a biosphere only (R0), or with a biosphere plus a technosphere (S5, S9, S10).

We can attempt to break this degeneracy by examining the spatial distribution of these three scenarios (Fig. 6). Scenario S5 includes an extended technosphere with poles at Earth and Mars, while S9 and S10 include a separation of the biosphere on Earth from the technosphere in the rest of the solar system. Further optical observations of Earth could be useful for identifying scenario S5, as this scenario has a high intensity of artificial illumination from urban areas (Fig. 7). However, the lack of artificial illumination would still be consistent with S9, S10, or R0.

The extended technospheres in S9 and S10 include activities such as mining in the asteroid and Kuiper belts, settlements across the outer solar system, and fusion propulsion, along with radio, optical, and quantum communication technosignatures (Table 8). Scenario S9 includes Dyson sphere elements, gravitational waves from super-luminal propulsion, and laser propulsion. These are all potential technosignatures that could theoretically be observed to conclude that the system has technosignatures (S9 or S10) or is a non-technological system (R0). But observing these technosignatures could be difficult, especially if communication leakage is transient, and Dyson sphere elements are diffuse. Observations of surface modification provide another option, as S9 includes near-complete surface modification of Mars and Venus, while S10 includes modification of the moon and Mars (Fig. 7). These features will also be difficult to resolve with near-term observing facilities, but such observations would be needed in this case to determine whether or not technology is present in the system.

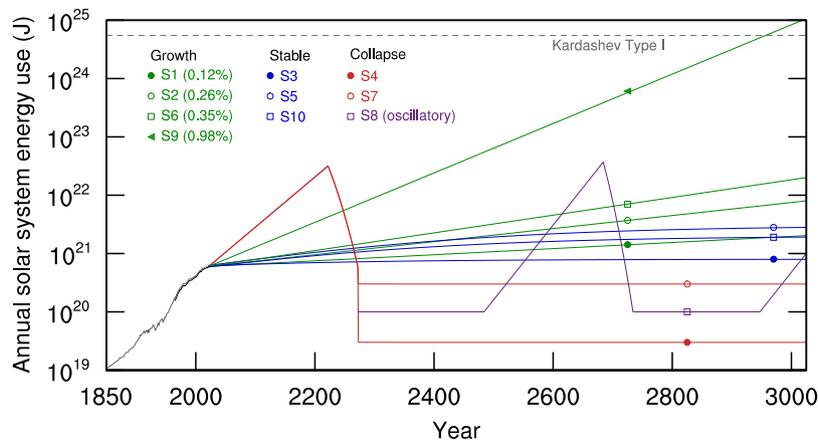

**Fig. 10.** Total annual energy use across all bodies in the solar system for each scenario. Approximate trajectories are shown for each scenario across the 1000-year timeline. Several scenarios have achieved stable (zero growth) conditions (S3, S5, S10), and one scenario (S8) is in a state of oscillation between growth and collapse. Only one (S9) is close to the 1% value assumed by Kardashev (1964), which has reached the Type I threshold by the end of the 1000-year period. Calculations assume that energy use is correlated with population at a value of 75 GJ per person per year (Jackson et al., 2022).

The S9 and S10 scenarios are particularly informative because Earth is part of the biosphere but not part of the technosphere. We can imagine a scenario in which a system analogous to S9 or S10 is discovered in the search for habitable exoplanets. Suppose that a future facility is able to measure visible and infrared spectra from the Earth-like planet in this system, which looks like the R0 spectra shown in Figs. 8 and 9. This spectra would have many of the spectral biosignatures that are the hallmark of life on Earth, so such an observation may be celebrated as the discovery of an exoplanetary biosignature. The spectra of this planet would show no absorption features that indicate industrial or agricultural pollution, nor would evidence of urban lighting be present. Many astrobiologists may conclude that this is an example of a planet with life, but no technology; however, such a conclusion would miss the fact that this system in fact has advanced technology spread across most of the solar system. Even attempts at spectral characterization of other planetary atmospheres would not necessarily reveal the presence of technology, and only later attempts at searching for surface modification would reveal that this in fact is a system teeming with technology.

In an effort to not miss the object of our search, it is important to remember that the relationship between biospheres and technospheres may not be obvious, and we should not expect the discovery of technosignatures (or biosignatures) to follow the patterns that we see in the solar system today or that we imagine from science fiction. The evidence of technology in an exoplanetary system may not necessarily be apparent in all cases from remote observation, and in-situ follow-up missions may be the only way to resolve any such degeneracies. This only serves to motivate further development of mission concepts that can advance the search for technosignatures (e.g., Socas-Navarro et al., 2021).

### 5.3. Longevity of technological civilizations

The most easily studied astrophysical phenomena are long-lived, whereas short-duration events are more difficult to observe. Such logic implies that the search for technosignatures will be more likely to discover evidence of long-lived technological civilizations rather than those with shorter lifetimes. This reasoning is captured by the Drake equation (Drake, 1965), which is a probabilistic expression for the number of detectable technosignatures in the galaxy, $N$. For our purposes, it is sufficient to write the Drake equation as $N \sim L$, where $L$ is the average communicative lifetime of technological civilizations. If $L$ is large, then the galaxy contains a larger fraction of systems with technosignatures at any given time, but if $L$ is small, then the galaxy contains only a few other extant civilizations—or perhaps we are the only one.

Our scenario set makes projections of Earth's 1000-year future; such future scenarios can be useful in guiding the search for technosignatures, but it is worth examining the extent to which we might expect 1000-year projections of the future to be representative of technospheres that might actually be targets of observation. The 1000-year projection is much greater than any other methodological approaches in futures studies, but a millennial timescale is still short by astronomical standards. We therefore complete our analysis of our scenarios in this study by considering the potential longevity of our scenarios as well as the extent to which we should expect long-lived civilizations to be prominent.

Three of our scenarios involve a civilization that has achieved stability, with zero net growth and long-term sustainability likely (S3, S5, S10). The S3 and S5 scenarios both include space tourism and industry out to Neptune, while S10 has a technosphere that extends across the entire solar system except Earth. None of these three scenarios involve a civilization that embarking on a program of interstellar expansion. Any interstellar activity in these scenarios remains limited to scientific exploration and for purposes other than expansion. These three scenarios could all conceivably extend out far beyond the 1000-year timeline and remain viable. These scenarios could also represent long-term steady-state conditions to the extent that no other major evolutionary or technological changes occur; because of this possibility, these three scenarios all remain possible candidates for thinking about the search for technosignatures. We also acknowledge the possibility that further evolutionary or technological changes could lead these scenarios to develop different technospheres over longer-term projections, and we return to this possibility once we complete our assessment of the other scenarios.

Two of our scenarios involve a civilization that has collapsed and has rebuilt itself in order to maintain long-term sustainability (S4, S7). The S4 and S7 scenarios both have technospheres limited to Earth only, with few technosignatures that could conceivably be detectable. These scenarios represent civilizations that maintain an environmental policy of minimizing impact on Earth, to the extent that the spectral signature of Earth in these scenarios is difficult to discern from a planet with no technosphere. Neither of these scenarios involve any spaceflight activity, let alone interstellar exploration. These scenarios represent long-term steady-state conditions and likewise remain candidates for thinking about the search for technosignatures. These scenarios may be difficult to detect, and it is worth considering the possibility that most long-lived technospheres in the galaxy may be those that are difficult to distinguish from biospheres.

One scenario involves a civilization that oscillates between growth and collapse, with the third collapse approaching (S8). The S8 scenario has a technosphere that extends out to the moon, but the technological capabilities of this civilization are limited to Apollo-era spaceflight. The long-term sustainability of this scenario is uncertain to the extent that the number of possible oscillations that could occur between growth and collapse is unknown. Further exploration of S8 over longer timescales could provide insight on whether this oscillation represents a steady-state or whether this oscillation will decay into a more permanent and stable condition.

Four of our scenarios involve a civilization that continues to grow (S1, S2, S6, S9). Scenario S1 has a low growth rate but is ultimately on a collapse trajectory due to the instability of its autocratic government. Further exploration of S1 over longer timescales would identify the extent to which a civilization could recover from such a collapse, but S1 otherwise is an example of a short-lived civilization. Scenario S2 includes space tourism and industry out to the asteroid belt, while S6 has space tourism and industry out to the Kuiper belt. The long-term sustainability of both of these scenarios is uncertain, and neither scenario includes any interest in interstellar exploration. Scenario S9 includes posthuman infrastructure that extends out to the Kuiper belt, with no imminent signs of collapse but also no guarantee of long-term sustainability. Scenario S9 involves the highest rate of growth among our scenarios and includes active interstellar exploration for eventual expansion. If these scenarios represent growth that is asymptotically approaching an equilibrium, then the technospheres in these scenarios may still be useful in thinking about the search for technosignatures. Nevertheless, exploration of all these scenarios over longer timescales could provide insight on the extent to which these represent stable or unstable futures.

Extending our worldbuilding pipeline to consider longer timescales remains conceptually possible to an extent. If we consider these future scenarios out to 10,000 years from now, our pipeline would need to consider the impact of geologic changes on Earth and other long-term physical processes on the biosphere and technosphere. Such a task is perhaps tractable with our existing methods, which may even be sufficient to decide whether our growth scenarios ever reach an equilibrium state. Further extensions out to 100,000 years from now would approach evolutionary timescales, which would require evaluating the evolutionary pressures that have selected for features of the biosphere and technosphere. Such work would require significant modification to the pipeline to enable consideration of the potential needs of future humans as well as the future evolution of the biosphere. Some prior work exists that could form the basis for such an effort, but this would be a significant undertaking. Even longer timescales of a million years or more are beyond the scope of this pipeline, as our approach based on human needs may not be relevant to such a long term future. Likewise, our scenario modeling eliminated many possibilities as inconsistent for our 1000-year scenarios, but these would all need to be revisited in the case of an extremely long timescale. Some scholars have speculated about the possible characteristics of Earth and human civilization one million years from now (Broderick, 2008), which represent multiple visions of the future but not necessarily self-consistent projections. These visions of a million-year future include speculations such as lifespans that "approach immortality", construction of a "Matrioshka brain" (star-sized computer), and the prediction that "everything will change but numbers and laughter" (Tonn, 2021). We do not make any assessments in this study about the likelihood of such outcomes, but we note that such long-term developments would be impossible to capture with our human-needs based methodology. We save exploration of 10,000-year and possibly 100,000-year futures for subsequent studies, but longer timescales will require the development of novel scenario modeling and worldbuilding methods.

As a final remark, it is possible that short-lived technosignatures are more abundant and the most likely to be detected. Rapid-growth scenarios like S9 may be rare or nonexistent, or such rapid-growth scenarios may inevitably collapse when evaluated on longer timescales. If this is the case, then shorter-lived technospheres will be those that are the most likely to be observed. This possibility was noted by Balbi and Grimaldi (2024), who suggested that "[if] short-lived technoemissions vastly outnumber the long-lived ones (as it is the case if their operation has an energy or maintenance cost that increases with time), then the first to be detected will likely have a relatively short $L$". We do not yet know whether long-lived technospheres are common or even possible, so we should not restrict the search for technosignatures to long-lived scenarios alone. If short-lived technospheres are the only ones that exist, then our set of 1000-year projections of Earth's future provides a rich set of possibilities for thinking about strategies for detecting these remote technospheres with explicit tracing of prior premises and assumptions.

## 5.4. Recommendations for technosignature detection

Further analysis of the scenario set developed in this study can be used to provide quantitative constraints on the search for technosignatures, such as by performing more sophisticated climate-chemistry modeling and then assessing the detectability of particular spectral features using instruments on current or future observatories. However, the results of the present analysis can still provide specific recommendations for advancing the search for technosignatures, in particular by re-examining concepts that are sometimes taken for granted in technosignature research. Three specific recommendations follow, based on the preceding discussion.

First is that increases in energy use and growth may not be tightly coupled, as assumed in many applications of the Kardashev scale, and instead a technological civilization may prioritize exploration of its spatial scale over exploitation of its available energy (e.g., Berger-Tal et al., 2014). Even our rapid growth scenario S9 has only reached the energetic limit equivalent to a Kardashev Type I civilization (i.e., using an energy equivalent to that available to Earth from incident solar radiation), but scenario S9 has a spatial scale that spans the entire solar system (which is the spatial scale associated with a Type II civilization). Technosignature searches that are predicated upon searching for Type I, II, or III civilizations in which the energetic and spatial scales are linked may find that no such civilizations exist at this precise limit, as all civilizations will require tradeoffs between prioritizing exploration and exploitation. Civilizations like Earth, and the scenarios considered in this study, may all be limited in their ability to harness luminous stellar energy and may not ever reach the theoretical limit of a Type I, II, or III civilization. Techosignature searches that seek to find evidence of a civilization that is approaching or even exceeding this limit would benefit from thinking more broadly about other possible metabolic mechanisms that prioritize exploitation over exploration. For example, the idea of a "stellivore" (Vidal, 2016) that harnesses energy by consuming stellar mass, rather than luminosity, is a theoretical possibility that would be observable as a compact accreting binary star system. Targeted searches of for such exotic technosignatures, such as analyses of compact accreting binaries (Vidal, 2019; Vidal et al., 2023; Vidal, 2024), provide an avenue for advancing technosignature searches beyond conventional approaches that look for Type I, II, or III civilizations.

Second is that spectral technosignatures may be degenerate or inconclusive when examining a single planet alone. The spectral characterization of terrestrial planets is becoming increasingly feasible and has motivated studies of spectral techosignatures that could conceivably be detectable with next-generation observatories (e.g., Kopparapu et al., 2021; Haqq-Misra et al., 2022a,b; Beatty, 2022); however, few, if any, of these spectral signatures would be conclusive evidence of extraterrestrial technology if observed in isolation. This is illustrated in our study with the degeneracy between the spectra of Earth in our technological scenarios (S5, S9, S10) and pre-agricultural Earth (R0). Breaking this degeneracy in our example would require searching for spectral technosignatures in the spectra of other planets (such as Mars or Venus), looking for artifact technosignatures in the outer parts of the system, or searching for communicative radio or laser signals. Targeted searches of planets of interest, such as planets thought to harbor liquid water or otherwise host habitable conditions, may be a necessary but insufficient condition for identifying the presence of a technosphere. In the best cases, the expansion of a technosphere to multiple planets in the system could provide corroborating evidence of extraterrestrial technology (such as a system that includes terraformed planets). In the worst cases, a planet of interest may harbor a technosphere that has no uniquely identifiable spectral signatures and no other detectable technosignatures. For such cases, exploration with remote spacecraft that can actually visit the planet may be the only way to detect evidence of such a technosphere. In general, efforts at spectral characterization of exoplanets should be conducted in tandem with other searches within the same system of interest in order to optimize the likelihood of finding evidence of a technosignature.

Third is that technospheres similar to present-day or near-future Earth could conceivably be the most abundant technospheres in the galaxy or universe and thus the most likely to be detected. This could be because most technospheres are short-lived and do not ever reach the point of sustaining themselves across geologic timescales, but this could also be because long-lived technospheres are not so drastically different from Earth today. Nine of our scenarios (all but S1) could conceivably be sustained across geologic timescales without significant changes to their technospheres. Using Earth as a benchmark for technosignature detection studies may be a reasonable assumption to begin with; even though technology may be a relatively recent development in the history of Earth, this does not necessarily imply that future developments will inevitably lead to drastic shifts in the detectability of Earth's technosphere. Our scenario set provides examples in which Earth's spectral signature in the future is not so different from that of today, and it remains possible that the most abundant or detectable extraterrestrial technosignatures are comparable in magnitude to those on Earth today.

## 6. Conclusion

This study is the first to generate self-consistent projections of Earth's technosphere for 1000 years into the future with an explicit tracing of prior premises and assumptions. Our scenario modeling approach builds upon existing methods in futures studies to define a scenario space and identify a smaller number of unique scenarios, which we tailored to generate a wide range of possible technospheres. Our worldbuilding pipeline was designed specifically for this study, which draws upon existing frameworks for assessing human needs and their satisfiers and defining the features of the technosphere. Our focus has remained specifically on understanding the technospheres of these civilization-level scenarios, but other applications may also find value in using of our scenario set to explore other features of these possible futures. Our novel worldbuilding pipeline may also have application in areas beyond the search for technosignatures, and we encourage other scholars to draw upon our method as a model or inspiration for systematically projecting deep futures that are based on the intertwinement of human systems and the technosphere.

Our overview of the remotely detectable technosignatures in this study highlights the diversity of technospheres that arise from our methodological approach. We save detailed assessment of detectability for subsequent work, but we show spectral absorption features for a few cases (Figs. 8 and 9) as an illustrative example of how these scenarios can inform technosignature search strategies. The $NO_2$ feature at visible wavelengths is a prominent way to distinguish between a present-day Earth scenario (R1), a pre-agricultural Earth scenario (R0), and an industrial future Earth scenario (S3). The differences between pre-agricultural and present-day Earth are significant enough to note that industry and agriculture are already affecting our spectral signature. And yet, the pre-agricultural Earth scenario itself is spectrally indistinguishable from scenarios S5, S9, and S10—all of which include technology across the solar system. This represents a viable possibility in the actual search for life in exoplanetary systems, and it is important to keep in mind the possibility that the technospheres that actually exist may not be the ones that are the easiest to observe.

The trajectories of these scenarios indicate the range of possibilities open to the future of human civilization. Our scenario set includes projections of future collapse, albeit not extinction-level ones, but also includes many examples of stable technospheres that have reached an equilibrium or have significantly reduced the rate of growth from today. These optimistic outcomes of slower- or zero-growth rates remain tenable and plausible trajectories that extend from today, even if they may not seem likely in context of present-day events. These scenarios also differ in the extent to which technology spreads to other parts of the solar system, which underscores the inherent uncertainty in the longevity of any present-day ambitions for the permanent settlement of space. We cannot predict which of our scenarios is more likely to resemble the actual future, but we can point to the existence of optimistic outcomes in our scenario set to show that our own civilization is not necessarily destined for collapse or extinction. The future remains open, and our collective efforts to envision possible futures should serve as a guide toward realizing the best actual future.

## CRediT authorship contribution statement

## Declaration of competing interest

The authors declare that they have no known competing financial interests or personal relationships that could have appeared to influence the work reported in this paper.

## Acknowledgments

The authors acknowledge support from the NASA Exobiology program, USA under grant 80NSSC22K1009. Thanks to Adam Frank, Connor Martini, and Pinchen Fan for helpful discussions. Any opinions, findings, and conclusions or recommendations expressed in this material are those of the authors and do not necessarily reflect the views of their employers or NASA.

## Appendix A. Supplementary data

Supplementary material related to this article can be found online at https://doi.org/10.1016/j.techfore.2025.124194.

## Data availability

The worldbuilding pipeline documents for each scenario are also available as supplementary material at https://dx.doi.org/10.5281/zenodo.11174443.

**Dr. Jacob Haqq Misra** is a research scientist an astrobiologist at the Blue Marble Space Institute of Science. His research focuses on understanding the conditions that allow life to survive on Earth and the possibility of detecting signatures of biology or technology on other planets. He also studies the possible futures of life in the solar system and is the author of the book *Sovereign Mars: Transforming Our Values through Space Settlement.*

**Dr. George Profitiliotis** is a research scientist and a futures studies expert at the Blue Marble Space Institute of Science. He has worked extensively in the fields of strategic foresight, futures literacy, and futures studies, and he is particularly interested in applying techniques from those fields to the domains of anticipatory governance and anticipatory ethics of advanced technoscience.

**Dr. Ravi Kopparapu** is a planetary scientist and an astrophysicist at NASA Goddard Space Flight Center. His research work includes identifying and characterizing habitable worlds in our galaxy, modeling the atmospheres of exoplanets, and exploring signatures of biology and technology on exoplanets. Dr. Kopparapu leads the Sellers Exoplanet Environment Collaboration (SEEC) at NASA Goddard and is science-PI of the CHAMPs team (Consortium on Habitability and Atmospheres of M-dwarf Planets).